\documentclass[12pt]{article}
\usepackage{amsmath}
\usepackage{epsfig}
\begin{document}
\title{Nucleon QCD sum rules in instanton medium
}
\author{M. G. Ryskin, E. G. Drukarev,~V. A. Sadovnikova\\
{\em National Research Center "Kurchatov Institute"}\\
{\em B. P. Konstantinov Petersburg Nuclear Physics Institute}\\
{\em Orlova Roscha, Gatchina, Leningrad district, 188300, Russia}}

\date{}

\maketitle

\begin{abstract}
We try to find the grounds for the standard nucleon QCD sum rules, based on more detailed description of QCD vacuum.
We calculate the polarization operator of the nucleon current in the instanton medium. The medium (QCD vacuum) is assumed to be a composition of the small-size instantons and some long-wave gluon fluctuations. We solve the corresponding QCD sum rules equations and demonstrate that there is a solution with the value of the nucleon mass close to the physical one if the fraction of the small size instantons contribution $w_s \approx 2/3$.
\end{abstract}

\section{Introduction}

The idea of the QCD sum rules approach is to express the characteristics of the observed hadrons in terms of the
vacuum expectation values of the QCD operators, often referred to as the condensates. It was suggested in \cite{1}
for calculation of the characteristics of mesons. Later it was used for the nucleons \cite{2}.
It succeeded in calculation of the nucleon mass, anomalous magnetic moment, axial coupling constant, etc \cite{3}.

The QCD sum rules (SR) approach is based on the dispersion relation for the
function describing the propagation of the system which carries the quantum
numbers of the hadron. This function is usually referred to as the "polarization operator"  $\Pi(q)$,
with $q$ the four-momentum of the system. The dispersion relation in which we do not take care of the subtractions
\begin{equation}
\Pi(q^2)=\frac1\pi\int dk^2\frac{\mbox{Im}\Pi(k^2)}{k^2-q^2}
\label{0}
\end{equation}
is analyzed at large and negative values of $q^2$. Due to the asymptotic freedom of QCD the polarization operator can be calculated in this domain. The Operator Product Expansion (OPE) \cite{4} enables to present the polarization operator as a power series
of $q^{-2}$ at $q^2\to-\infty$. The coefficients of the expansion are the QCD condensates, such as the scalar quark condensate $\langle 0|\bar q(0)q(0)|0\rangle$, gluon condensate $\langle 0|G^{a\mu \nu}G^a_{\mu \nu}|0\rangle$, etc. The nonperturbative physics is contained in these
condensates. The typical values of condensate with the dimension $d=n$ is $\langle 0|O_n|0\rangle \sim (\pm 250 {\rm MeV})^n$. Thus  we expect the series $\Pi(q)=\sum_n\langle 0|O_n|0\rangle/(q^2)^n$ to converge at $-q^2 \sim 1~\,{\rm GeV}^2$.

The left hand side (LHS) of Eq.~(\ref{0}) is calculated as the OPE series. The imaginary part on the right hand side (RHS) describes the physical states with the baryon quantum number and charge equal to unity. These are the proton, described by the pole of ${\rm Im} \Pi(k^2)$, the cuts corresponding to the systems containing the proton and pions, etc.
The right-hand side  of Eq.~(\ref{0}) is usually approximated by the "pole+continuum" model
\cite{1,2} in which the lowest lying pole is written down exactly,
while the higher states are described by continuum.
The main aim is to obtain the value of the nucleon mass.

The polarization operator can be written as
\begin{equation}
\Pi(q^2)=i\int d^4xe^{i(q\cdot x)} \langle 0|T[j(x)\bar j(0)]|0 \rangle\,,
\label{6}
\end{equation}
where $j(x)$ the local operator with the proton quantum numbers,
often referred to as ``current". It is a composition of the quark operators.
Thus the integrand on the RHS of Eq.~(\ref{6}) contains the nonlocal expectation values
$\langle 0|\bar q(0)q(x)|0\rangle$.
The nonlocal condensates have been considered earlier (see \cite{3a} and references therein), mainly in the studies of the pion wave functions.

Note that the product $\bar q(0)q(x)$ is not gauge invariant. This expression makes sense if we define $q(x)$ as the expansion near the point $x=0$, i.e.,
\begin{equation}
q(x)=\,\left(1+x^{\mu}D_{\mu}+\frac{x^{\mu}x^{\nu}}{2}D_{\mu}D_{\nu}+...\right) q(0)\,,
\label{300a}
\end{equation}
with $D^{\mu}$ standing for covariant derivatives. One usually employs the Fock-Schwinger (fixed point) gauge with $x_{\mu}A^{\mu}(x)=0$ where $A^{\mu}(x)$ is the gluon field. This enables to replace the covariant derivatives by the ordinary ones. Thus the condensate $\langle 0|\bar q(0)q(x)|0\rangle$ can be expressed as a Taylor series, containing the set of new condensates, such as $\langle 0|\bar q(0)\partial^2q(0)|0\rangle$.

In this approach the QCD condensates are considered as phenomenological parameters. Extractions of their  values from experimental data, supported by certain theoretical ideas does not always lead to unique conclusions.

One usually applies the Borel transform which converts the functions of $q^2$ to the functions
of the Borel mass $M^2$. Note also that the Borel transform removes the divergent contributions caused by behavior of the integrand in the integral on the RHS of Eq.~(\ref{6}) at $x \rightarrow 0$. An important assumption is that there
is an interval of the values of $M^2$ where the two sides of the SR have a good
overlap, approximating also the true functions. This interval is in the region of $1\,{\rm GeV}^2$. Thus actually
one tries to expand the OPE from the high momentum region to the region of $ |q^2| \sim 1\,{\rm GeV}^2$.

To calculate the polarization operator defined by Eq.~(\ref{6}) we must clarify the form of the current $j(x)$.
 It is not unique. One can write
\begin{equation}
j(t,x)=j_1(x)+tj_2(x)\,,
\label{7}
\end{equation}
with
$$j_1(x)=(u^T_a(x)Cd_b(x))\gamma_5u_c(x) \varepsilon^{abc}, \quad j_2(x)=(u^T_a(x)C\gamma_5d_b(x))u_c(x) \varepsilon^{abc},$$
while $t$ is an arbitrary coefficient. Following \cite{9} we use the current determined by Eq.~(\ref{7})
with $t=-1$, which can be written (up to a factor $1/2$) as
\begin{equation}
j(x)=(u^T_a(x)C\gamma_{\mu}u_b(x))\gamma_5 \gamma^{\mu}d_c(x) \varepsilon^{abc}\,.
\label{8} \end{equation}
This current is often used in the QCD SR calculations. One of the strong points of the choice is that it makes the domination of the lowest pole over the higher states on the RHS of the SR more pronounced. We use only this current in the present paper.

Any model of QCD vacuum should explain the origin and the values of QCD condensates.
A popular nowadays point of view (see, e.g., \cite{5}) is that the QCD vacuum is filled by strong gluon fields (instantons). The values of the QCD condensates are determined by the space-time structure of the instantons.
Hence, the instantons provide more detailed description of vacuum than the QCD condensates.

We try to write the QCD sum rules in terms of the parameters of the instanton vacuum. Our aim is not to replace the OPE approach, but to study a possible role of a more detailed structure of the QCD vacuum.

The instanton medium is characterized by the distribution of the instantons over their sizes $\rho$ which we denote as $n(\rho)$ and by the distances between the instantons $R$ which also have certain  distribution. The distribution $n(\rho)$
is known to peak at $\rho \approx 0.33$ fm \cite{5}. The paper \cite{6} presents the summary of a number of lattice calculations of the distribution. Detailed analysis of the distribution over sizes is given in \cite{6a}. As to the distance between the instantons, the conventional assumption is that the averaged separation is $R \approx 1$ fm. \cite{5}.
We employ the simplest model which reproduces the essential physics of the process.
We
assume that the QCD vacuum consists of the "small-size" instantons
with $\rho_s \approx 0.33$ fm (we shall variate this value) and
  some long wave gluon field fluctuations characterized by scale $\rho_{\ell} \gg 1$GeV${^-1}$.
Thus the quarks composing the polarization operator move in superposition of the fields of the small size instantons and some long wave fluctuations.


We treat the quarks in the field of small instantons, following the approach developed in \cite{7,8}.
In this approach the light quarks move in the self-consistent field of interacting small-size instantons. They are described by the propagator (in Eucledian metric)
\begin{equation}
S_{ab}(p)=\frac{\hat p+im(p)}{p^2+m^2(p)}\delta_{ab}\,,
\label{31a}
\end{equation}
with the effective dynamical mass $m(p)$ found in \cite{7,8}.  Note that the value of $R=1$~fm allows to reproduce in such an instanton vacuum model the value of the gluon condensate. This instanton-instanton separation $R=1$~fm is much larger than the
inverse Borel mass $1/M \sim 0.2$~fm. Thus the size of the system described by the polarization operator (\ref{6}) is much smaller
than $R$ and can accommodate only one ("nearest") instanton. Recall that the latter is a part of self-consistent system of interacting instantons.  This leads to several consequences.
We can write the quark propagator in the "nearest instanton approximation" (NIA) rather as
\begin{equation}
S_{ab}(p)=S_Z+S_{NZ}; \quad S_Z(p)=\frac{im(p)}{p^2}\delta_{ab}; \quad S_{NZ}(p)=\frac{\hat p}{p^2}\delta_{ab}\,.
\label{31}
\end{equation}
Here $S_Z$ is the zero-mode contribution. The sum of all the nonzero-mode contributions $S_{NZ}$ is approximated thus by the free propagator of the massless quark. Note that in the NIA we include only the terms which are proportional to the instanton density.

In this approach one of the nonvanishing contributions
 comes from the configurations where all quarks are described by propagators $S_{NZ}$. Another contribution  comes from the configuration where the $u$ quarks are described by $S_{NZ}$, i.e., do not feel the instantons while the $d$ quark is described by the propagator $S_Z$.
The other configurations do not contribute to the polarization operator since two $u$ quarks can not be in zero mode of the same instanton. The configuration in which $u$ and $d$ quarks are in the zero modes correspond to the $SU(2)$ version of the instanton induced 't Hooft interaction \cite{tH}. For the current (\ref{8}) this configuration does not contribute since it contains the trace of odd (three) number of $\gamma$ matrices which turns to zero.

Comparing now the structures of chirality conserving and chirality flipping components  of polarization operator
in condensate and small-size nearest instantons "languages" we see that they differ. In both "languages" the chirality conserving structure contains the loop of three free quarks. However, there is no such thing as the four-quark condensate in the "nearest instanton approximation" \footnote{Going beyond the terms which are linear in the instanton density, we would obtain configuration with two $u$ quarks in the instanton field. In the limit $M^2\rho_s^2 \rightarrow  \infty$ the contribution corresponds to the factorized four-quark condensate in the OPE language-see below.}. In the nucleon QCD sum rules $M^2$ is of the order $1$ GeV$^2$, and thus $\tau =M^2\rho_s^2  \sim 1$. On the other hand, the quark condensate created by the small size instantons can be represented by the general relation
\begin{equation}
\langle 0|\bar q(0)q(0)|0\rangle_s=i\int\frac{d^4p}{(2\pi)^4}{\rm Tr} S(p) = -4N_c \int \frac{d^4p}{(2\pi)^4} \frac {m(p)}{p^2}\,
\label{8a} \end{equation}
(lower index $s$ means the small size instantons, $N_c$ is the number of colors), creating a bridge between the instanton and condensate "languages".
In the limit $\tau \gg 1$ the two languages provide the same result, and the contribution is proportional to the
quark condensate $\langle 0|\bar q(0)q(0)|0\rangle$.
In the instanton picture at $\tau \sim 1$ the contribution  can be viewed as coming from the nonlocal scalar condensate $\langle 0|\bar q(0)q(x)|0\rangle_s$. The nonlocal condensate is not a new subject, it was employed earlier for example in the pion QCD sum rules in \cite{3a}.

The radiative corrections containing the terms $\alpha_s\ln{M^2}$ (the leading logarithmic approximation, LLA) to the chirality conserving structure are the same as in the OPE case. The same refers to  LLA corrections to the chirality
flipping structure, since they originate from the $u$ quarks loop and are determined by large momenta of the virtual gluons, which  exceed strongly the momentum
carried by the current.

We demonstrate that the QCD sum rules constructed in such a way do not have a physical solution. Thus we must assume that the small instantons create only part of the scalar condensate. From the first glance this contradicts the results of \cite{7}, \cite{8},  where the small instantons reproduced the conventional value of
$\langle 0|\bar q(0)q(0)|0\rangle$. However, in these papers the instanton density, which is proportional to $1/R^4$ is tied to the gluon condensate which is known with large uncertainties (up to a factor $2$) \cite{IR}. This leaves some room for other contributions to the quark scalar condensate. Here we assume that the small-size instantons provide a fraction $w_s$ of the total scalar condensate. Our model assumption is that the rest part $(1-w_s)\langle 0|\bar q(0)q(0)|0\rangle$ is due to interactions at a large correlation length $\rho_l \gg 1$GeV$^{-1}$. It can be approximated by a local condensate.
Thus the chirality flipping component of the polarization operator is determined by terms which describe interactions of the $d$ quark with the nearest small-size instanton and by a local condensate. We write for  the expectation value
\begin{equation}
\langle 0|\bar q(0)q(x)|0\rangle=\langle 0|\bar q(0)q(x)|0\rangle_{s}+\langle 0|\bar q(0)q(0)|0\rangle_{\ell}; \quad
\langle 0|\bar q(0)q(x)|0\rangle_{s,\ell}=\langle 0|\bar q(0)q(x)|0\rangle w_{s,\ell};
\label{6a0}
\end{equation}
$$w_s+w_{\ell}=1,$$
where we did not account for nonlocalities of the second term. This realizes the old idea \cite{F1}, \cite{F2} that the large-size instanton contributions are included in condensates, while the small-size instantons provide nonperturbative contributions writen explicitely. As a special case  the condensate $\langle 0|\bar q(0)q(0)|0\rangle_{\ell}$ can be treated as due to the long-size instantons with $\rho_{\ell}M \gg 1$

Now the polarization operator obtains a contribution from configuration in which one of the $u$ quarks moves in the zero mode of the small-size instanton, while the second one is described by a local scalar condensate. In the limit
$\tau \gg 1$ the leading term of expansion in powers of $1/\tau$  is
 equal to that given by the standard total condensate of the OPE approach.

For $M^2$ of the order of $1$ GeV$^2$ we have $\rho_s^2M^2\sim 1$, and the convergence of the OPE series is obscure. For the chirality flipping sum rule the function of $M^2$ on the LHS can be viewed as
coming from the nonlocality of the scalar quark condensate.
The contribution of the four-quark condensate presented in instanton picture provides now a much smaller contribution since one instanton can produce only one $\bar qq$ pair of the fixed flavor. On the other hand some of the condensates which contribute to the OPE SR are not accounted for in our model
where all the nonzero modes contribution is approximated (included) by the free quark propagator $S_{NZ}$.

We calculate the polarization operator $\Pi(q)$ in the instanton vacuum in the NIA and analyze the corresponding SR.
We demonstrate that the SR have a solution with the value of the nucleon mass not far from the physical one for
all $w_s < 0.6-0.7$. At $w_s \approx 2/3 $ the value of the nucleon mass is $m_N \approx 1$ GeV. Comparing with the SR in terms of condensates we found that the consistency between the LHS and RHS of the SR improved. At the conventional values of the quark condensate $\langle 0|\bar q(0)q(0)|0\rangle \approx (-250 $MeV$)^3$ the value of the threshold does not change much, while that of nucleon residue becomes noticeably smaller.
At larger values of $w_s$ the SR have only an unphysical solution with the contribution of the continuum exceeding much that of the nucleon pole.
In Sec.~2 we recall the main features of the nucleon  SR in terms of condensates. In Sec.~3 we calculate the polarization operator in instanton vacuum. In Sec.~4 we solve the SR equations. We discuss the results in Sec.~5.

\section{QCD sum rules in terms of condensates}

In the case of nucleon (we consider the proton) the polarization operator takes the form
\begin{equation}
\Pi(q)=\hat q\Pi^q(q^2)+I\Pi^I(q^2)\,,
\label{1}
\end{equation}
with $q$ the four-momentum of the system, $\hat q=q_{\mu}\gamma^{\mu}$, $I$ is the unit matrix. The first and the second terms on the RHS correspond to the chirality conserving and the chirality flipping contributions, correspondingly. The dispersion relations are
 \begin{equation}
\Pi^i(q^2)=\frac1\pi\int dk^2\frac{\mbox{Im}\Pi^i(k^2)}{k^2-q^2}\,;
 \quad i=q,I .
\label{2}
\end{equation}
As we said earlier, we do not take care of the subtractions.

We present the results of calculation of  the polarization operator defined by Eq.~(\ref{6}) with the current determined by (\ref{8}).
The LHS of Eq.~(\ref{2}) can be written as
\begin{equation}
\Pi^{q~OPE}(q^2)=\sum_{n=0}A_n(q^2); \quad \Pi^{I~OPE}(q^2)=\sum_{n=3} B_n(q^2)
\label{3a}
\end{equation}
where the lower index $n$ is the dimension of the corresponding QCD condensate ($A_0$ stands for the three-quark loop).
The most important terms for $n\leq 8$ were obtained earlier \cite{2,3}.
For the chirality conserving structure they are
\begin{equation}
A_0=\frac{-Q^4\ln{Q^2}}{64\pi^4}; \quad A_4=\frac{-b\ln{Q^2}}{128\pi^4}; \quad A_6=\frac{1}{24\pi^4}\frac{a^2}{Q^2};
\quad A_8=-\frac{1}{6\pi^4}\frac{m^2_0a^2}{Q^4}\,.
\label{3c}
\end{equation}
Here $Q^2=-q^2>0$, while $a$ and $b$ are the scalar and gluon condensates multiplied by certain numerical factors
\begin{equation}
a=-(2\pi)^2\langle 0|\bar q q|0\rangle  ; \quad b=(2\pi)^2\langle 0|\frac{\alpha_s}{\pi}G^{a\mu \nu}G^a_{\mu \nu}|0\rangle\,,
\label{3d}
\end{equation}
while
\begin{equation}
m^2_0 \equiv \frac{\langle 0|\bar q\sigma_{\mu\nu}{\cal G}_{\mu\nu}q|0\rangle}{\langle 0|\bar q q|0\rangle};
\quad {\cal G}_{\mu\nu}=\frac{\alpha_s}{\pi}\sum_hG^h_{\mu\nu}\lambda^h /2\,.
\label{3dd}
\end{equation}
We shall discuss the value of the $m_0^2$ in Sec.~5.
For the chirality flipping structure we find
\begin{equation}
B_3=\frac{aQ^2\ln{Q^2}}{16\pi^4}; \quad B_5=0\,.
\label{3ddd}
\end{equation}
The leading contribution to the chirality conserving structure $ A_0$ is the loop containing three free quarks. The leading contribution to the chirality-odd structure $B_3$ is proportional to the scalar quark condensate. Here the free $u$ quarks form a loop, while the $d$ quarks are exchanged with the vacuum condensate - see Fig.~1.

Let us, however, pay attention to the latter equality $B_5=0$. There are indeed two contributions of dimension
$d=5$. Thus we can write $B_5=B_5^a+B_5^b$. The term $B_5^a$ comes from the Taylor expansion of the product
$\bar d(0)d(x)$ and is proportional to the condensate $\langle 0|\bar d(0)D^2d(0)|0\rangle$. In this case
the $u$ quarks are described by free propagators which are diagonal in color variables. However the product
of the operators $d^a_{\alpha}\bar d^b_{\beta}G^h_{\mu\nu}$ provide the  contribution to the propagator of $d$ quark which is proportional to the product $\langle 0|\bar q {\cal G}_{\mu\nu}\sigma_{\mu\nu}q|0\rangle\cdot\sigma_{\alpha\beta}\lambda^h_{ab}/2$. The contribution to polarization operator $B^b_5$ is thus
proportional to the condensate $\langle 0|\bar q{\cal G}_{\mu\nu}\sigma_{\mu\nu}q|0\rangle$, and the propagator
of one of the $u$ quarks of polarization operator should include interaction with this gluon field (and can not be treated as a free one). Due to equation of motion
$$ (D^2-\frac{1}{2}\sigma_{\mu\nu}{\cal G}_{\mu\nu})q=m^2_qq,$$
with $m_q$ standing for the current mass of the quark, one finds that for the massless quark
\begin{equation}
\langle 0|\bar d(0)D^2d(0)|0\rangle=\frac{1}{2}\langle 0|\bar q\sigma_{\mu\nu}{\cal G}_{\mu\nu}q|0\rangle\,.
\label{4ddd}
\end{equation}
Thus the contributions $B_5^a$ and $B_5^b$  can be expressed in terms of the same condensate.
Direct calculation \cite{I1} demonstrates that $B_5^a+B^b_5=0$. Note that this cancelation takes place only for the current (\ref{8}). If one employs the current (\ref{7}) with $t \neq -1$, the contribution $B_5 \neq 0$.

Actually one usually considers the SR for operators ${\cal P}^i(M^2)=32\pi^4{\cal B}\Pi^{i\,OPE}(q^2)$ with ${\cal B}$
the operator of Borel transform. The factor $32\pi^4$ is
introduced in order to deal with the values of the order of unity (in
GeV units). After the Borel transform we write (\ref{3a}) as
\begin{equation}
{\cal P}^q(M^2)=\sum_{n=0}A'_n(M^2); \quad {\cal P}^i(M^2)=\sum_{n=3} B'_n(M^2);
\quad A'_n(M^2)=32\pi^4{\cal B}A_n(q^2)\,.
\label{3b}
\end{equation}
$$ B'_n(M^2)=32\pi^4{\cal B}B_n(q^2).$$
Here we present the most important terms
\begin{equation}
A'_0(M^2)=M^6; \quad A'_4(M^2)=\frac{bM^2}{4}; \quad
A'_6=\frac43 a^2;\quad
B'_3(M^2)=2aM^4\,.
\label{22a}
\end{equation}
The Borel transformed SR (\ref{2}) can be written now as
\begin{equation}
{\cal P}^i(M^2)={\cal F}_p^{i}(M^2)+{\cal F}_c^{i}(M^2)\,,
\label{3g}
\end{equation}
where the two terms on the RHS are the contributions of the nucleon pole with the mass $m_N$ and that of the continuum
\begin{equation}
{\cal F}_p^{i}(M^2)=\xi_i\lambda^2e^{-m_N^2/M^2};
\quad {\cal F}_c^{i}(M^2)=\int_{W^2}^{\infty}dk^2e^{-k^2/M^2}\Delta[{\cal B}\Pi_1(k^2)]
\label{3h}
\end{equation}
to the RHS of the Borel transformed Eq.~(\ref{2}). Here $\lambda^2$ is the residue at the nucleon pole (multiplied by $32\pi^4$), $W^2$ is the continuum threshold;
$\xi_q=1$, $\xi_I=m_N$.

The conventional form of the SR is
\begin{equation}
{\cal L}^q(M^2, W^2)=R^q(M^2)\,,
\label{4}
\end{equation}
and
\begin{equation}
{\cal L}^I(M^2, W^2)=R^I(M^2)\,.
\label{4a}
\end{equation}
Here ${\cal L}^i$ and $R^i$ are the Borel transforms of the LHS and of the RHS of Eqs.~(\ref{2}), correspondingly
\begin{equation}
R^q(M^2)=\lambda^2e^{-m_N^2/M^2}; \quad R^I(M^2)=m_N\lambda^2e^{-m_N^2/M^2}\,,
\label{5}
\end{equation}
with $\lambda^2=32\pi^4\lambda_N^2$. The contribution of continuum is moved to the LHS of
Eqs.~(\ref{4}, \ref{4a}) which can be written as
\begin{equation}
{\cal L}^q=\sum_{n=0} \tilde A_n(M^2, W^2); \quad {\cal L}^I=\sum_{n=3}\tilde B_n(M^2, W^2)\,,
\label{5a}
\end{equation}
-see Eq.~(\ref{3b}). Here

\begin{equation}
\tilde A_0=\frac{M^6E_2(\gamma)}{L(M^2)};\quad
\tilde A_4=\frac{bM^2E_0
(\gamma)}{4L(M^2)};
\label{22}
\end{equation}
$$\tilde A_6=\frac43 a^2L; \quad
\tilde B_3=2aM^4E_1(\gamma); \quad \gamma=\frac{W^2}{M^2}\,,$$
with
\begin{equation}
E_0(\gamma)=1-e^{-\gamma}, \quad E_1(\gamma)=1-(1+\gamma)e^{-\gamma}, \quad E_2(\gamma)=1-(1+\gamma+\gamma^2/2)e^{-\gamma}\,.
\label{23} \end{equation}
The factor
\begin{equation}
L(M^2)=\Big(\frac{\ln M^2/\Lambda^2}{\ln \mu^2/\Lambda^2}\Big)^{4/9}
\label{3100} \end{equation}
includes the most important radiative corrections of the order $\alpha_s\ln{Q^2}$ (LLA). These contributions were summed to all orders of $(\alpha_s\ln{Q^2})^n$.
In Eq.~(\ref{3100}) $\Lambda=\Lambda_{QCD}$ is the QCD scale, while $\mu$ is the
normalization point, the standard choice is $\mu=0.5\,$GeV.

The position of the nucleon pole $m_N$, its residue $\lambda^2$ and the continuum threshold $W^2$ are the unknowns of the SR equations (\ref{4}) and (\ref{4a}). The nucleon sum rules equations (\ref{4}) and (\ref{4a}) are usually solved at $M^2 \sim 1\,{\rm GeV}^2$, namely
\begin{equation}
0.8 \mbox{GeV}^2 \leq M^2 \leq 1.4\,\mbox{GeV}^2\,.
\label{230}
\end{equation}
The interval of the values of $M^2$ where the SR are true is usually referred to as "duality interval".

After inclusion of several condensates of the higher dimension and of the lowest order radiative corrections beyond the leading logarithmic approximation \cite{10}
the SR provide solution (for $\Lambda_{QCD}=230$ MeV) $m_N=928$ MeV, $\lambda^2=2.36\,{\rm GeV}^6$, $W^2=2.13\,{\rm GeV}^2$.

\section{QCD sum rules in instanton vacuum}
\subsection{Instanton presentation and OPE approximation}

Recall that a typical value of the condensate of dimension $d=n$ is $\langle 0|O_n|0\rangle \sim c^n$ with $c=250\,$MeV. Since at $M \approx 1$ GeV we had $c^2M^2 \ll 1$, we could expect the convergence of the OPE. In the instanton language we have $\rho_s^2M^2 \sim 1$ and can not expect the convergence of the series in powers of $1/M^2$.

Thus the structure of the LHS of the sum rules differs from that in the condensate presentation. The leading contribution $A_0$ to the chirality conserving operator $\Pi^q$ remained unchanged. However, while we consider only the nearest
small-size instanton, there is no contribution of two zero-mode $u$ quarks (this contribution plays the role of the four-quark condensate in the "condensate language"), since only one $u$ quark can be placed in the zero mode
of the field of the instanton.

In the chirality flipping structure $\Pi^I$ we describe the $d$ quark by the propagator $S_Z$ given by Eq.~(\ref{31}).
The Borel transformed contribution ${\cal B}\Pi^I(M^2)$ depends on the parameter $\rho_s^2M^2 \sim 1$ and can not be presented as $1/M^2$ series. In the condensate language this means that it includes the nonlocal scalar quark condensate.

Note that our form for the propagator $S_{NZ}$ means that we did not pick some of contributions which were present in the condensate picture. In the terms $A_4$ and $B_5^b$ the propagator of one of the $u$ quarks should
include the influence of the gluon field. Thus its propagator is not diagonal in color indices, while both $S_Z$ and $S_{NZ}$ are.

Now we assume that in the NIA the small-size instantons create only the part $w_s<1$ of the scalar condensate. The contribution of  the rest part of the condensate $(1-w_s)a$ to the chirality flipping structure
is expressed by the term $B_3$ of Eq.(\ref{22a}) with $a$ replaced by $a_{\ell}$. In the chirality flipping structure
one of the $u$ quarks is in the zero-mode of the nearest small-size instanton, while the other one
is described by a local condensate. The latter provides  the factor $a_{\ell}$ in  the contribution to the polarization operator. The former one provides a factor containing the nonlocal scalar condensate.
Note that here we did not use the factorization hypothesis.

Considering only the small-size instantons, we do not have an analog of the OPE four-quark condensates since
two $u$ quarks can not be in the zero-mode of the same instanton. We find such analog going beyond the NIA.

\subsection{Calculation of polarization operator}

Instead of the condensate $\langle 0|\bar q(0)q(0)|0\rangle$ we shall employ parameter $a$ defined by Eq.(\ref{3d}).
We introduce also $a_s=aw_s$ and $a_{\ell}=aw_{\ell}=a(1-w_s)$. It these variables Eq.(\ref{8a}) with $N_c=3$  takes the form.
\begin{equation}
a_s=6\int_0^{\infty}dp\,p\,m(p); \quad a_s=-(2\pi)^2\langle 0|\bar q q|0\rangle_s\,.
\label{33A} \end{equation}

As we said earlier, the leading contribution  $A_0$ to the $\hat Q$ structure
remains unchanged. The contribution to the chirality flipping structure is now
\begin{equation}
\Pi_1^I(q^2)=2a(1-w_{s})Q^2\ln{Q^2}+Y_s\,,
\label{34}
\end{equation}
where the two terms are the contributions of the large size and small size instantons, correspondingly. The last one can be written as $Y_s=32\pi^4X_s$, with
\begin{equation}
X_s=12\int\frac{d^4p}{(2\pi)^4}\gamma_{\mu}\frac{m(p)}{p^2}\gamma_{\nu}T_{\mu\nu}(Q-p)\,,
\label{48}
\end{equation}
while
\begin{equation}
T_{\mu\nu}(Q-p)=\int d^4x e^{-i(Q-p,x)}{\rm Tr}[t_{\mu\nu}(x)]\,,
\label{35}
\end{equation}
with
\begin{equation}
t_{\mu\nu}(x)=\gamma_{\mu}G_0(x)\gamma_{\nu}G_0(x)\,.
\label{36}
\end{equation}
Here
\begin{equation}
G_0(x)=-\frac{1}{2\pi^2}\frac{\hat x}{x^4}
\label{37}
\end{equation}
is the Fourier transform of the propagator $S_{NZ}$ determined by Eq.~(30).
 Note that now the quark in zero mode carries a nonzero momentum $p$. In the condensate language it carries momentum $p=0$.
Neglecting momentum $p$ in the factor $ T_{\mu\nu}(Q-p)$ on the LHS of Eq.~(\ref{48}) we would obtain
\begin{equation}
X_s=\frac{3Q^2\ln{Q^2}}{8\pi^4}\int_0^{\infty} dp\,p\,m(p)=B_3(Q^2)\,,
\label{50}
\end{equation}
with $B_3(Q^2)$ defined by Eq.~(\ref{3ddd}) and $a$ replaced by $a_s$.
Thus in the limit $Q^2 \rightarrow \infty$ we obtain the lowest OPE term.
We can view calculation of the contribution given by Eq.~(\ref{48}) as inclusion of nonlocality in the scalar quark condensate.

The four-quark contribution can emerge only if one of the $\bar u u$ pairs comes from the small size instantons, while the other one originates from large size fluctuation. Following the previous discussion we can write the contribution to polarization operator as
\begin{equation}
A_6=\frac{4aw_s(1-w_s)}{\pi^2}\int \frac{d^4p}{(2\pi)^4}\frac{m(p)}{p^2}
\frac{\hat Q-\hat p}{(Q-p)^2}\,.
\label{63}
\end{equation}
Here the lower index $6$ shows that neglecting $p$ in the last factor on the RHS we would obtain the factorized OPE term
$\tilde A_6$ determined by Eq.~(\ref{22a}) multiplied by $2w_s(1-w_s)$.
The set of diagrams included in the SR is shown in Fig.~2.

It is instructive to trace how the contributions to the spin-conserving part of the polarization operator change if we
we go beyond the NIA. In the loop corresponding to $A_0$ the quark propagators have now the masses squared in denominators, and the contribution of $A_0$ diminishes. Now two $u$ quarks can be described by the chirality flipping parts of their propagators. The corresponding contribution of the small-size instantons is
\begin{equation}
A'_6=12\cdot8\cdot\int \frac{d^4p}{(2\pi)^4}\frac{m(p)}{p^2}\int \frac{d^4p'}{(2\pi)^4}\frac{m(p')}{p'^2}
\frac{\hat Q-\hat p-\hat p'}{(Q-p-p')^2}\,.
\label{63'}
\end{equation}
In the limit $Q^2 \rightarrow \infty$ we neglect $p$ and $p'$ in the last factor of the integrand, coming to the factorized form of the OPE contribution.

To obtain results in analytical form we parameterize the dynamical quark mass caused by the small size instantons
\begin{equation}
m(p)=\frac{{\cal A}}{(p^2+\eta^2)^3}\,,
\label{53}
\end{equation}
with ${\cal A}$ and $\eta$ the adjusting parameters. The power of denominator insures the proper behavior
$m(p) \sim p^{-6}$ at $p \rightarrow \infty$ \cite{7}.
Now Eq.~(\ref{33A}) can be written as
\begin{equation}
a_s=\frac{3{\cal A}}{2\eta^4}\,.
\label{54}
\end{equation}

Calculating the tensor $T_{\mu\nu}$ we present
\begin{equation}
X_s=\frac{3}{\pi^2}\int \frac{d^4p}{(2\pi)^4} \frac{{\cal A}}{p^2(p^2+\eta^2)^3}(Q-p)^2\ln{(Q-p)^2}\,.
\label{49}
\end{equation}

Further details of calculations are presented in Appendix. We find for the Borel transformed
contribution
\begin{equation}
B'(M^2)=2a_{\ell}M^4+2a_sM^4F(\beta); \quad F(\beta)=\frac{1}{3}\Big(\frac{2(1-e^{-\beta})}{\beta}+e^{-\beta}(1-\beta) +\beta^2{\cal E}(\beta)\Big)\,;
\label{55}
\end{equation}
$$\beta=\eta^2/M^2.$$
Here
\begin{equation}
{\cal E}(\beta)=\int_{\beta}^{\infty}dt\frac{e^{-t}}{t}\,.
\label{56}
\end{equation}
In literature our function ${\cal E}$ is usually denoted as $E_1$. We avoid this notation since in QCD SR
publications the notation $E_1$ has another meaning - see Eq.~(\ref{23}).

Combining Eq.~(\ref{54}) with the relation $m(0)={\cal A}/\eta^6$ coming from Eq.~(\ref{53}) we find that $\eta^2=2a_s/3m(0)$. It was demonstrated in \cite{7,8} that $a_s \sim R^{-2}\rho^{-1}$, while $m(0)\sim
R^{-2}\rho$. Thus $\eta^2$ depends only on $\rho$, and  $\eta^2=1.26\,{\rm GeV}^2$ at $\rho=0.33$~fm.
In the duality interval (\ref{230}) $0.9 \leq \beta \leq 1.6$.
The function $F(\beta)$ is shown in Fig.~3a.
The dependence of the function $F$ on $M^2$ for $\eta^2=1.26\,{\rm GeV}^2$ is shown in Fig.~3b.
As expected, in the limit $M^2 \rightarrow \infty$ we find $B=B'_3$ with the latter defined by Eq.~(\ref{22a}).

Similar calculation provides
\begin{equation}
A'_6=\frac{8}{3}a^2w_s(1-w_{s})\frac{1-e^{-\beta}}{\beta}\,.
\label{63a}
\end{equation}

\subsection{Parametrization of the nonlocal scalar condensate}
It is reasonable to try to establish connection with the OPE approach. We write Eq.~(\ref{55}) as
\begin{equation}
B'_3(M^2)=2M^4a(M^2)\,,
\label{80cc}
\end{equation}
where
\begin{equation}
a(M^2) = a\left(1-w_s+w_sF(\frac{\eta^2}{M^2})\right)\,,
\label{80ca}
\end{equation}
while $F$ is defined by Eq.~(\ref{55}). We have $a(M^2) \rightarrow a$ at $M^2 \rightarrow \infty $.
Now we define
\begin{equation}
K(M^2)=\frac{a(M^2)}{a}\,,
\label{80c}
\end{equation}
and try to parametrize the function $K(M^2)$ in the duality interval as a power series in $1/M^2$:
\begin{equation}
K(M^2)=1+\sum_{n=1}^{N}C_n/M^{2n}\,,
\label{80d}
\end{equation}

If the second term on the RHS can be approximated by one or two terms,
such presentation can be related to the parametrization
of the expectation value
$\langle 0|\bar q(0)q(x)|0\rangle$ by a polynomial in $x^2$.
We can write the polarization operator $\Pi_I$ as
\begin{equation}
\Pi_I(q^2)=\frac{2}{\pi^4}\int \frac{d^4x}{x^6}f(x^2)e^{iqx}\,,
\label{80g}
\end{equation}
with $f(x^2)=\langle 0|\bar q(0)q(x)|0\rangle$.
Assuming that $f(x)$ can be parameterized as
(recall that we are in Euclidean metric)
\begin{equation}
f(x)=f(0)(1+c_1x^2+c_2x^4)\,,
\label{81xx}
\end{equation}
we find
\begin{equation}
B'_3(M^2) = 2M^4f(0)\left(1-\frac{8c_1}{M^2}+\frac{32c_2}{M^4}\right)\,,
\label{81yy}
\end{equation}
and thus in (\ref{80d})
\begin{equation}
C_1=8c_1; \quad C_2=32c_2\,.
\label{81yyy}
\end{equation}
Note that the RHS of Eq.~(\ref{81xx}) can not be treated as the lowest terms of the Taylor expansion. The terms $x^{2n}$ with $n \geq 3$ provide the integrals which are divergent on the upper limit and can not be eliminated by the Borel transform.

For the medium consisting solely of the small size instantons, i.e., for $w_s=1$ keeping the first three terms in (\ref{80d}) we find that
$C_1=-1.23\mbox{ GeV}^2; \quad C_2=0.54\mbox{ GeV}^4\,$
and thus $c_1=-0.15\,{\rm GeV}^2$, $c_2=0.017\,{\rm GeV}^4$ in the duality interval interval $0.8\,{\rm GeV}^2 \leq M^2\leq 1.4\,{\rm GeV}^2$ determined by Eq.~(\ref{230}).
The accuracy of the parametrization is illustrated by Fig.~4.
In the interval $0.8\,{\rm GeV}^2 \leq M^2\leq 2.0\,{\rm GeV}^2$
we find slightly different set of values
$C_1=-1.38\mbox{ GeV}^2; \quad C_2=0.68\mbox{ GeV}^4\,$
corresponding to $c_1=-0.17\,{\rm GeV}^2$, $c_2=0.021\,{\rm GeV}^4.$
Thus we can assume that parametrization (\ref{81xx}) with $c_1 \approx -0.2\,{\rm GeV}^2$, $c_2\approx 0.02\,{\rm GeV}^4.$ can be employed for the Borel masses is $GeV$ region.
This point was discussed in more details in \cite{M1}.

\section{Solution of the sum rules equations}

Now we return to the Minkowsky metric and analyze  Eqs.~(\ref{4}) and (\ref{4a}) with
\begin{equation}
{\cal L}^q=\tilde A_0(M^2, W^2)+ \tilde A_6(M^2)  ; \quad {\cal L}^I=\tilde B(M^2, W^2)\,.
\label{67a}
\end{equation}
Here $\tilde A_0(M^2, W^2)$ is given by Eq.~(\ref{22}), $\tilde A_6=A'_6$ is presented by Eq.~(\ref{63a}), while
\begin{equation}
\tilde B(M^2,W^2)=2a_{\ell}M^4E_2(\gamma)+2a_sM^4\Phi(M^2, W^2)\,;
\label{67}
\end{equation}
$$\Phi(M^2, W^2)=\frac{1}{3}\Big(\frac{2}{\beta}(1-e^{-\beta})+e^{-\beta}(1-\beta)-e^{-\gamma}(1-\beta+\gamma) +\beta^2({\cal E}(\beta)- {\cal E}(\gamma)) \Big)\,.$$
The functions $E_i (i=0,1,2)$ are determined by Eq.~(\ref{23}).

\subsection{Lack of solution at $w_s=1$}

One can guess immediately that there is no solution for $w_s=1$.
 Indeed, if the values $m_N, \lambda^2, W^2$ compose a solution, one
should obtain
\begin{equation}
\kappa(M^2)\equiv \frac{{\cal L}^I(M^2,W^2)}{{\cal L}^q(M^2,W^2)} \approx const=m_N
\label{167}
\end{equation}
since the contribution of the continuum should not be too large, we should expect
\begin{equation}
\frac{{\cal L}^I(M^2)}{{\cal L}^q(M^2)} \approx const \approx m_N\,,
\label{167a}
\end{equation}
where we put ${\cal L}^i(M^2)={\cal L}^i(M^2,W^2 \rightarrow \infty)$.

For $w_s=1$ Eq.~(\ref{167a}) takes the form
\begin{equation}
\kappa(M^2)=\frac{2aF(\eta^2/M^2)}{M^2}\,.
\label{167b}
\end{equation}

Employing the dependence of the function $F$ on $M^2$ for $\eta^2=1.26\,{\rm GeV}^2$ presented in Fig.~3b,
one can see that the values of $\kappa$ varies  between $2a\cdot 0.36/{\rm GeV}^2$ and $2a\cdot 0.27/{\rm GeV}^2$
in the interval (\ref{230}) of variation of $M^2$. For the distance $R=1$~fm between the small size instantons
$a=0.59\,{\rm GeV}^3$ \cite{7,8}.
Thus we obtained $m _N\approx 0.35$ GeV.

The unrealistic value of the nucleon mass obtained in such a way is, however, not the main problem.
Let us try to find the value of $\lambda^2$ employing Eq.~(21). We obtain $M^6\,e^{m^2_N/M^2} = \lambda^2$.
However the LHS of this equality changes by a factor of $6$ in the duality interval (\ref{230}).
Thus it can be satisfied only if the contribution of the continuum, which as it was discussed around Eqs.~(\ref{4} - \ref{5a}) was moved to the LHS, changes the LHS strongly.
Hence, we come to an unphysical solution of the sum rules \cite{11a}. As we shall see below, a more detailed analysis confirms this conclusion.

Note that at $w_s=1$ the RHS of SR for both chirality flipping and chirality conserving structures suffered large changes
comparing to the standard OPE SR. The most important change in the former case is inclusion of the nonlocality of the scalar condensate. In the latter case there is no four-quark condensate, which played important role in the OPE case.

\subsection{Dependence of the solutions on the fraction of small size instantons}
The functions ${\cal L}^q$ and ${\cal L}^I$ depend explicitly on the scalar condensate $a$, on its fraction caused by the
instantons of the small size $a_s=aw_s$ and on the parameter $\eta^2$. On the other hand, the medium of the small instantons is determined by their average size $\rho_s$ and the distance between the instantons $R(w_s)$. It was found in
\cite{7,8} that
\begin{equation}
\langle 0|\bar q(0)q(0)|0\rangle_s =\frac{C}{R^2(w_s)\rho}_s\,,
\label{67b}
\end{equation}
with $C=25.0$, $R$ is the distance between the small-size instantons. Thus we can study dependence of the solution of the sum rules equations on the fraction of the small size instantons $w_s$ for several values of the scalar condensate $a=-(2\pi)^2 \langle 0|\bar q(0)q(0)|0\rangle$ and
for  different sizes of small instantons $\rho_s$.

Note that at $\rho_s=0.33$~fm and $R(1)=1$~fm the scalar condensate $a=0.58\,{\rm GeV}^3$ (at the conventional normalization point $\mu=0.5$ GeV) \cite{7,8}.
This enables us to find the dependence on $w_s$ at any values of $a$ and $\rho_s$.

The results for $\rho_s=0.33$~fm are presented in Table 1 and in Fig.~5.  One can see that at several reasonable values of the quark condensate the sum rules have a physical solution for $w_s$ which does not exceed certain value
$w_0$. At $w_s=w_0 \approx 0.67$ the solutions jump to unphysical ones with a smaller value of the nucleon mass and the dominative contribution of the continuum \cite{11a}. At $w_s$ about $0.6$ the nucleon mass is close to the physical value.

In Table 1 and in Fig.~5 we present the results for four values of the scalar condensate $a$ corresponding to
$\rho_s=0.33$~fm and the distances between the small instantons $R=1.3$~fm, $R=1.2$~fm,  $R=1.1$~fm and $R=1.0$~fm at $w_s=0.6$.
The distances $R=1.3$~fm and $R=1.2$~fm correspond to the values $a=0.58\,{\rm GeV}^3$ and $a=0.67\,{\rm GeV}^3$, i.e., to the values of the scalar condensate $\langle 0|\bar q(0)q(0)|0\rangle$ equal to $(-244\,{\rm MeV})^3$ and  $(-257\,{\rm  MeV})^3$, close to conventional values.
The distances $R=1.1$~fm and $R=1.0$~fm correspond to  $a=0.80\,{\rm GeV}^3$ and $a=0.96\,{\rm GeV}^3$, i.e., to somewhat larger values of $\langle 0|\bar q(0)q(0)|0\rangle$ equal to $(-273\,{\rm MeV})^3$ and to less realistic $(-290\,{\rm MeV})^3$.
The consistency of the LHS and RHS of the sum rules is illustrated by Fig.~6.

As we said earlier, the pole-to-continuum ratio
\begin{equation}
r_i(M^2)={\cal F}_i^{p}(M^2)/{\cal F}_i^{c}(M^2); \quad i=q,I
\label{80}
\end{equation}
of the two contributions to the RHS of Eq.~(\ref{3g}) is the characteristic of validity of the "pole + continuum" model for the spectrum of polarization operator - Eqs.~(\ref{3g},\ref {3h}). For larger values of $r_i(M^2)$ the model is justified better. The values of the ratio are presented in Table 2 for  $\rho_s=0.33$ ~fm, $w_s=0.60$. We took two cases for illustration. For $a=0.58\,{\rm GeV}^3$ solution is
\begin{equation}
m_N = 1.01\,{\rm GeV}; \quad \lambda^2=1.2\,{\rm GeV}^6; \quad W^2=2.0 \mbox{GeV}^2\,.
\label{80a}
\end{equation}
The pole-to-continuum ratio decreases with the value of $M^2$ - see Table 2.
Although
the SR equations can be solved with good accuracy in the broad interval of the values of the Borel mass (see Table 3),
the pole-to-continuum ratio becomes unacceptably small for $M^2>1.4\,{\rm GeV}^2$.
Thus in this case we stay in the traditional duality interval determined by Eq.~(\ref{230}).

For the condensate $a=0.96~$GeV$^3$, corresponding to $R(0.6)=1$~fm the solution is
\begin{equation}
m_N = 1.15\,{\rm GeV}; \quad \lambda^2=2.8\,{\rm GeV}^6; \quad W^2=2.9~{\rm GeV}^2\,.
\label{80z}
\end{equation}
Here the SR equations also can be solved with good accuracy in the large interval of the values of the Borel mass - see Table 3.
One can see that both $r_q$ and $r_I$ decrease while value of $M^2$ increases.
In this case the pole-to-continuum ratio is much larger than it was for the smaller values of the condensate.
Thus the interval of the values of $M^2$ where the SR equations can be solved becomes larger.

We also fix the value of $R=1.3$~fm and  trace the dependence of the solutions on variation of $\rho_s$. In Table 4 we present the results for $\rho=0.25$~fm ($a=0.76\,{\rm GeV}^3$; $\langle 0|\bar q(0)q(0)|0\rangle=(-268\,{\rm MeV})^3)$ and $\rho_s=0.40$~fm
($a = 0.48\,{\rm GeV}^3$; $\langle 0|\bar q(0)q(0)|0\rangle = (-230\,{\rm MeV})^3)$.
They are shown in Fig.~7. The situation is similar to the previous case when we changed $R$. However, at $\rho_s=0.40$~fm the jump to the unphysical solution takes place at a larger value
$w_s \approx 0.75$.

For $w_s=0.65$ the function $K(M^2)$ determined by Eq.~(\ref{80c}) is approximated by the series in the RHS
of Eq.~(\ref{80d}) with parameters
\begin{equation}
C_1=-0.80\mbox{ GeV}^2; \quad C_2=0.35\mbox{ GeV}^4\,,
\label{90}
\end{equation}
and thus $c_1=-0.10\,{\rm GeV}^2$, $c_2=0.011\,{\rm GeV}^4$.

\section{Summary}

We calculated the polarization operator of the nucleon current in the instanton medium which we assumed to be a composition of the instantons of  small size and of some large-size gluon field fluctuations with the correlation length $\rho_{\ell} \gg 1$ GeV$^{-1}$. The instantons of large size $\rho \gg  (1\,{\rm GeV})^{-1}$ manifest themselves in terms of the local scalar quark condensate. The quark propagator in the field of the small size instantons contained the zero mode chirality flipping part proportional to effective quark mass $m(p)$ and the nonzero mode part approximated by the propagator of the free massless quark \cite{7,8}. The zero-mode part can be expressed in terms of the nonlocal scalar condensate.

We solved the sum rules equations and trace the dependence on the solution on the fraction of the small size instantons $w_s$. We demonstrated that at $w_s \leq 0.6-0.7$ the sum rules have a solution with a reasonable value of the nucleon mass.
At $w_s \approx 2/3$ the value of the nucleon mass is very close to the physical one. The numerical values vary
slightly with variation of the actual values of the size of small instantons and of the distance between them.
Finally at the values of the scalar condensate close to the conventional value $(-250\,{\rm MeV})^3$
\begin{equation}
m_N \approx 1\mbox{ GeV}; \quad \lambda^2 \approx 1\mbox{ GeV}^6; \quad W^2 \approx 2\mbox{ GeV}^2\,.
\label{101}
\end{equation}
At larger values of $w_s$ the sum rules have only an unphysical solution with a strong domination of the continuum contribution over that of the nucleon pole and with a small value of the nucleon mass.

The solution (\ref{101}) was found for $\rho_s=0.33$~fm, while $R=1.2-1.3$~fm. It is valid also for $R\approx 1.3$~fm while $\rho_s \sim 0.25-0.40$~fm. Note that in \cite{7},\cite{8} the value of $R$ is tied to that of the gluon condensate. The latter is known with large uncertainty \cite{IR}, and $R=1.2$ fm is not unrealistic. Also, (see \cite{6a})
one can tie the gluon condensate  to the total instanton density. For conventional value $\langle 0|\frac{\alpha_s}{\pi}G^{a\mu \nu}G^a_{\mu \nu}|0\rangle/32\pi^2=(200 $MeV$)^4$ and the distance between the small-size instantons $R=1.2$ fm the densities of the small-size and large-size instantons are approximately the same.

At larger values of the quark condensate the values of the nucleon residue and of the continuum threshold increase, reaching the values $\lambda^2 \approx 3\,{\rm GeV}^6$ and $W^2\approx 3\,{\rm GeV}^2$ at $\langle 0|\bar q(0)q(0)|0\rangle=(-290\,{\rm MeV})^3$.

Comparing to the SR in the condensate presentation, we included the nonlocality of the scalar condensate.
Also the instanton presentation strongly diminished the role of the contribution corresponding to four quark condensate in the condensate language.

The consistency between the LHS and RHS of the sum rules appeared to be much better than in the sum rules
in terms of local condensates, where the value of "$\chi^2$ per point" was of the order $10^{-1}$ \cite{10} assuming the 10\% error bars. The mean relative difference between the LHS and RHS  is about 3\%. At larger values of the scalar condensate the domination of the contribution of the pole over that of the continuum becomes more pronounced. Also, the duality interval becomes larger than that defined by Eq.~(\ref{230}) due to the shift of the upper limit.

We demonstrated that the contribution of nonlocality of the  scalar condensate can be approximated by two additional terms of $1/M^2$ series. This corresponds to approximation of the dependence of the nonlocal quark condensate $f(x^2)=\langle 0|\bar q(x)q(0)|0\rangle$ on $x^2$ by a polynomial of the second order. At $x^2=1\,{\rm GeV}^{-2}$ (Euclidean metric) we found $f(x^2)-f(0)=tf(0)$ with $t=-0.14$ for $w_s=1$ and $t=-0.09$ for $w_s=0.65$. More complicated calculations in framework of the instanton liquid model \cite{Shur} provided $t \approx -0.1$ for $x^2=1\,{\rm GeV}^{-2}$.
The parameter $m_0^2$ defined by Eq.~(\ref{3dd}) determines the lowest order term of the Taylor series of the condensate $f(x^2)$. Its value was estimated in the nucleon QCD sum rules analysis as providing the best fit of the two sides of the sum rules.
The result of \cite{I1} is $m_0^2 \approx 0.8\,{\rm GeV}^2$ leading to $t \approx 0.2$, while the value  $m_0^2 \approx 0.2\,{\rm GeV}^2$ providing $t \approx-0.05$ was obtained in \cite{1a}.

Note that these are to large extent the preliminary results. Presenting the continuous distribution of the instanton sizes  as the superposition of instantons of  small size and of some large-size gluon field fluctuations we neglected their possible interactions. Another point is the interpretation of the condensate $(1-w_s) \langle 0|\bar qq|0\rangle$
caused by interactions at the large scale. A  more general analysis should be carried out. The last but not the least, we plan to include interactions between the quarks composing the polarization operator, i.e., we must take into account the radiative corrections. They are the same as in the condensate presentation for the structure $\Pi^q$. However, additional work is required to find these corrections for the chirality flipping structure $\Pi^I$. Thus, a more general analysis is required.
The results will be published elsewhere.

We thank A.E. Dorokhov, N. I. Kochelev and especially V. Yu. Petrov for stimulating discussions.

\def\thesection{Appendix \Alph{section}}
\def\theequation{\Alph{section}.\arabic{equation}}
\setcounter{section}{0}

\section{}
\setcounter{equation}{0}


In order to calculate the integral on the RHS of Eq.~(\ref{49}) we present
\begin{equation}
\ln{(Q-p)^2}=-\int_0^{\infty}\frac{dy}{(Q-p)^2+y}\,.
\label{A2}
\end{equation}
Here and below we omit the polynomials in $Q^2$ since they will be eliminated by the Borel transform.
Now we can write
\begin{equation}
X_s=-\frac{3i}{\pi^2}\int\frac{d^4p}{(2\pi)^4} \frac{{\cal A}}{p^2(p^2+\eta^2)^3} \int_0^{\infty}\frac{dy\,y}{(Q-p)^2+y}\,.
\label{A3}
\end{equation}
One can check that it is possible to present
\begin{equation}
\frac{1}{p^2(p^2+\eta^2)^3}=3\int_0^{1}\frac{dx\,x^2}{(p^2+\eta^2x)^4}\,,
\label{A4}
\end{equation}
and thus
\begin{equation}
X_s=-3\int_0^{1}dx\,x^2\Psi(\eta^2x)\,,
\label{A5}
\end{equation}
where
\begin{equation}
\Psi(\mu^2)=\frac{3{\cal A}}{\pi^2}\int_0^{\infty}dy\,y\Phi(\mu^2,y); \quad \Phi(\mu^2,y)=\int \frac{d^4p}{(2\pi)^4} \frac{1}{(p^2+\mu^2)^4}\cdot \frac{1}{(Q-p)^2+y}\,.
\label{A6}
\end{equation}
Carrying out the integration over the angular variables we find
\begin{equation}
\Phi(\mu^2,y)=\frac{1}{48\pi^2}\int_0^1\frac{dt(1-t)^3}{(ty+\mu^2(1-t)+t(1-t)Q^2)^3}=\int_0^1\frac{dt(1-t)^3}{t^3(y+\kappa)^3}; \quad \kappa=Q^2(1-t)+\frac{\mu^2(1-t)}{t^3}\,.
\label{A7}
\end{equation}
Carrying out integration over $y$ we obtain
\begin{equation}
\Psi(\mu^2)=\frac{{\cal A}}{32\pi^4}\int_1^{\infty}{du}\left(1-\frac{1}{u}\right)^2\frac{u}{Q^2+\mu^2u}\,.
\label{A8}
\end{equation}
The divergence on the upper limit is not important, since this contribution will be eliminated by the Borel transform.
Returning to Eq.~(\ref{A5}) we can write it as
\begin{equation}
X_s=\frac{3{\cal A}}{32\pi^4}\int_0^1dx\,x^2\int_1^{\infty}{du} \left(1-\frac{1}{u}\right)^2 \frac{u}{Q^2+\eta^2ux}\,.
\label{A9}
\end{equation}
Now integration can be carried out easily, providing
\begin{equation}
X_s=\frac{3{\cal A}}{32\pi^4}\Big[\frac{Q^4}{\eta^6}\ln{\frac{Q^2+\eta^2}{Q^2}}+(\frac{3Q^2}{\eta^4}+\frac{3}{\eta^2}+
\frac{1}{Q^2})\ln{\frac{Q^2+\eta^2}{\eta^2}}\Big]\,.
\label{A10}
\end{equation}
After the Borel transform we come to Eq.~({\ref{55}).
Note that carrying ont the Borel transform of the RHS of Eq.~(\ref{A9}) we come to a compact expression
\begin{equation}
{\cal B}X_s=\frac{3{\cal A}}{32\pi^4}\int_0^1dx\,x^2\int_1^{\infty}du\,u \left(1-\frac{1}{u}\right)^2 exp(-\eta^2xu/M^2)\,.
\label{A11}
\end{equation}

{}

\clearpage

\newpage
\begin{table}
\caption{Solutions of the sum rules equations for $\rho=0.33$ fm.}
\begin{center}
\begin{tabular}
{|c|c|c|c|c|c|} \hline
$a$,\, GeV$^3$&$w_s$&$m_N,$\,GeV&$\lambda^2$,\,GeV$^6$&$W^2$,\,GeV$^2$&$\chi^2_N\cdot 10^2$\\
\hline
      &0.30&1.45 & 8.7&6.6&3.7\\
0.96 &0.60&1.15 & 2.8 &2.9&4.0\\
      &0.66&1.05 & 1.9 &2.3&3.9\\
      &0.67&0.82  & 0.86&1.4 &2.1\\

\hline

      &0.30&1.40 &6.2 &4.9 &1.7\\
0.80 &0.60&1.10 &2.0  &2.6 &2.3\\
      &0.67&0.99  &1.2  &2.0 &2.2\\
      &0.68&0.80  &0.60 &1.3 &1.2\\

\hline

      &0.30 &1.33& 4.3  &4.0 &0.83\\
0.67 &0.60 &1.05& 1.4  &2.2 &1.4\\
      &0.67 &0.95 & 0.83 &1.7 &1.3\\
      &0.68 &0.77 & 0.41 &1.1 &0.59\\

\hline
      &0.30& 1.27& 3.0 & 3.4 &4.2\\
0.57 &0.60& 1.00& 0.95& 1.9 &0.89\\
      &0.67& 0.90 & 0.57& 1.5 &0.83\\
      &0.68& 0.75 & 0.30& 1.0 &0.30\\

\hline

\end{tabular} \end{center}
\end{table}

\begin{table}
\caption{Pole-to-continuum ratio $r(M^2)$ for solutions of the sum rules at $\rho=0.33$ fm
for $a= 0.58\,{\rm GeV}^3$ and $a= 0.96\,{\rm GeV}^3$; $w_s=0.60$.}

\begin{center}
\begin{tabular}{|c|c|c|c|} \hline
$a,\,{\rm GeV}^3$ & $M^2,\,{\rm GeV}^2$ & $r_q(M^2)$ & $r_I(M^2)$ \\
\hline

&0.8&1.25&1.84\\

0.58 &1.0&0.69&1.08 \\

&1.2&0.43&0.72 \\

&1.4&0.29&0.52\\

\hline
      &0.8& 4.69& 5.85\\
0.96 &1.0& 2.30& 2.99 \\
      &1.2& 1.34& 1.82 \\
      &1.4& 0.86& 1.23\\

\hline

\end{tabular} \end{center}
\end{table}

\begin{table}
\caption{Solutions of the sum rules equations in various intervals of the values of the Borel mass.
The values of parameters are the same as in Table 2.}
\begin{center}
\begin{tabular}{|c|c|c|c|c|c|} \hline
$a,\,{\rm GeV}^3$ & $M^2,\,{\rm GeV}^2$ & $m_N,\,{\rm GeV}$ & $\lambda^2,\,{\rm GeV}^6$ & $W^2,\,{\rm GeV}^2$ & $\chi^2_N\cdot 10^2$\\
\hline

      &0.8 - 1.4&1.01&0.98&1.96&0.93 \\

0.58 &0.8 - 1.6&1.02&1.01&1.99&1.2\\

      &0.8 - 1.8&1.03&1.04&2.01&1.5\\
\hline

      &0.8 - 1.4& 1.15& 2.83& 2.93 & 4.0 \\
0.96 &0.8 - 1.6& 1.17& 3.03& 3.02 & 5.1\\
      &0.8 - 1.8& 1.19& 3.20& 3.08 & 6.0\\
\hline
\end{tabular} \end{center}

\end{table}

\begin{table}
\caption{Solutions of the sum rules equations for $R \approx 1.3$ fm.}

\begin{center}
\begin{tabular}
{|c|c|c|c|c|c|} \hline
$a,\,{\rm GeV}^3$ & $w_s$ & $m_N$,~GeV & $\lambda^2,\,{\rm GeV}^6$ & $W^2,\,{\rm GeV}^2$ &
$\chi^2_N\cdot 10^2$\\
\hline

      &0.60 & 1.09& 1.52 & 2.31 &0.38\\
0.77 &0.70 & 0.88 & 0.55 & 1.38 &0.15\\
      \hline
      &0.50 & 1.06& 1.20 & 2.21 & 0.78\\
0.48 & 0.60&0.98 &0.82   &1.87 & 1.1\\
      \hline

\end{tabular} \end{center}
\end{table}

\clearpage

\section*{Figure captions}
\noindent
Fig.~1. The set of the diagrams for the lowest OPE terms of the nucleon sum rules.  Wavy lines are for the nucleon current, solid lines stand for the quarks, dashed lines denote the gluons. The circles stand for the quark and gluon condensates.\\

\noindent
Fig.~2. The set of the diagrams for the  quarks in the fields of instantons.
 Dark and dashed blobs on the quark lines stand for the small size and large size instantons.\\

\noindent
Fig.~3.~$a$: The function $F(\beta)$ determined by Eq.~(41).\\
$b$: Dependence of the  functions $F(\eta^2/M^2)$ for $\eta^2=1.26\,{\rm GeV}^2$, corresponding to the size $\rho_s=0.33$~ fm.\\
\noindent
Fig.~4. Approximation of the function $K(M^2)$ defined by Eq.~(\ref{80c}) (solid line) by the series on the RHS of Eq.~(\ref{80d}) with parameters determined by Eq.~(\ref{81yyy}) (dotted).

\noindent
Fig.~5. Dependence of the solution of the sum rules equations on the value of $w_s$ at $\rho_s=0.33$~fm. Fig.~5$a$ is for the nucleon mass, Fig.~5$b$ demonstrates the $w_s$ dependence of $\lambda^2$, Fig.~5$c$ is for $W^2$. The solid, dashed, dotted and dashed-dotted lines are for the values of the scalar condensate
$a=0.58\,{\rm GeV}^3$, $a=0.67\,{\rm GeV}^3$,  $a=0.80\,{\rm GeV}^3$, and $a=0.96\,{\rm GeV}^3$.

Fig.~6. Consistency of the $LHS$ and $RHS$ of the sum rules for the case $a=0.58\,{\rm GeV}^3$, $w_s=0.60$.
The solid and dashed lines show the RHS-to-LHS ratios for the SR for chirality conserving and chirality flipping equations, correspondingly. \\

\noindent
Fig.~7. Dependence of the solution of the sum rules equations on the value of $w_s$ at $R \approx 1.3$~fm for Fig.~7$a$ is for the nucleon mass, Fig.~7$b$ demonstrates the $w_s$ dependence of $\lambda^2$, Fig.~7$c$ is for $W^2$. The solid, dashed and dotted  curves are for the values of the scalar condensate $a=0.58\,{\rm GeV}^3$, $a=0.48\,{\rm GeV}^3$ and $a=0.77\,{\rm GeV}^3$,  correspondingly.\\

\clearpage
\newpage
\begin{figure}
\centerline{\epsfig{file=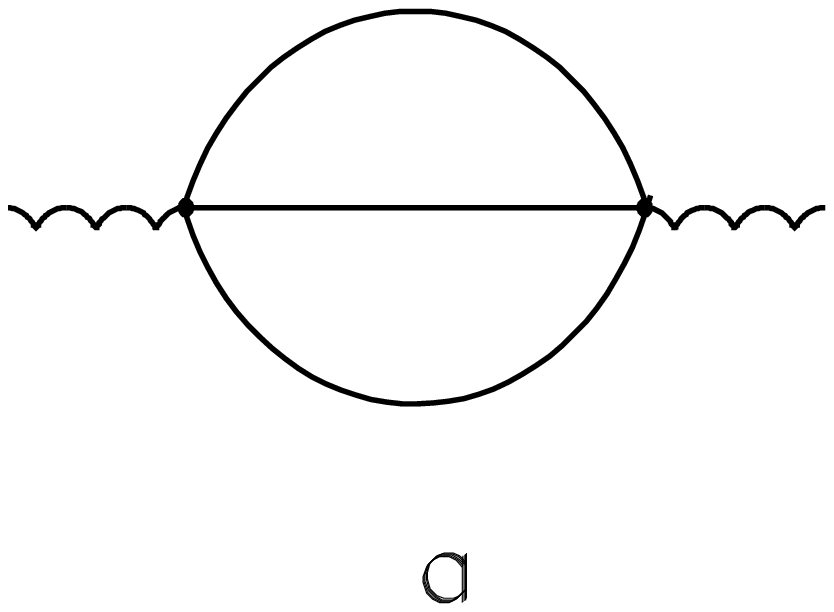,width=5cm}
\epsfig{file=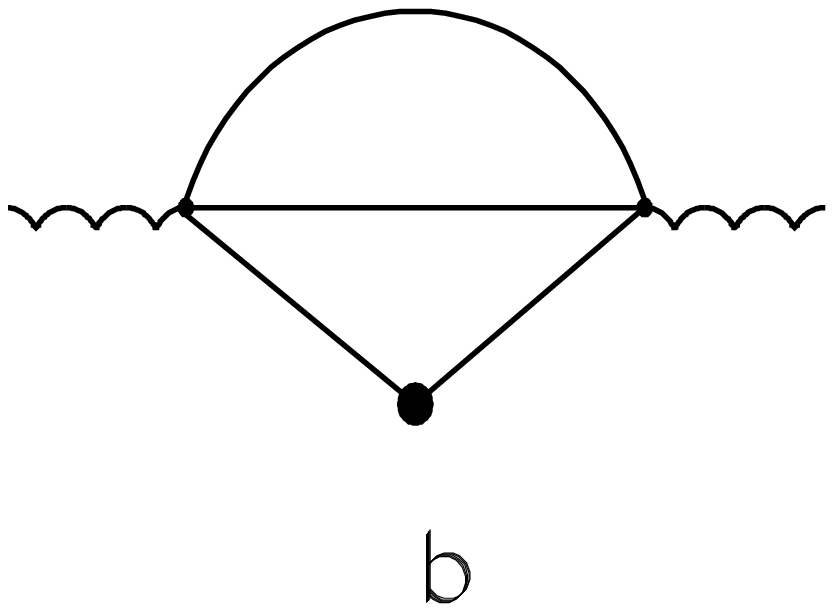,width=5cm}}
\centerline{\epsfig{file=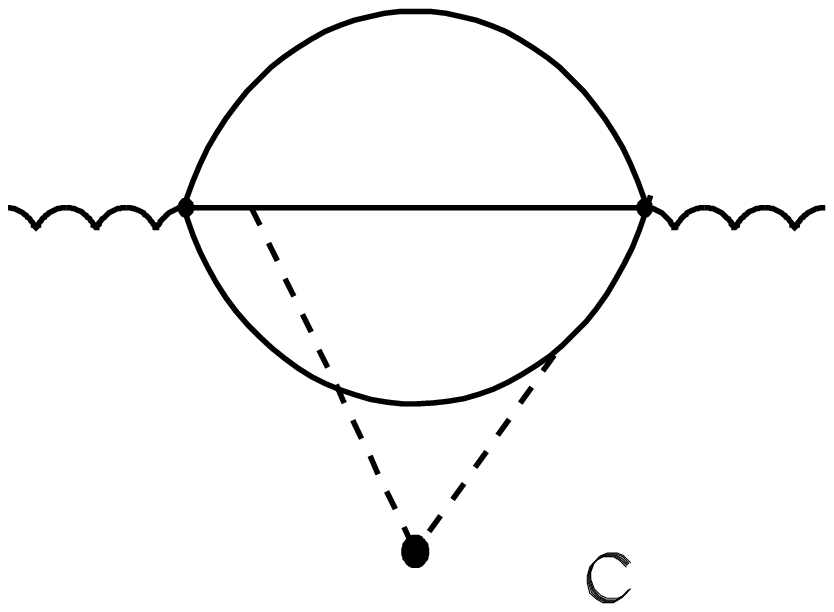,width=5cm}
\epsfig{file=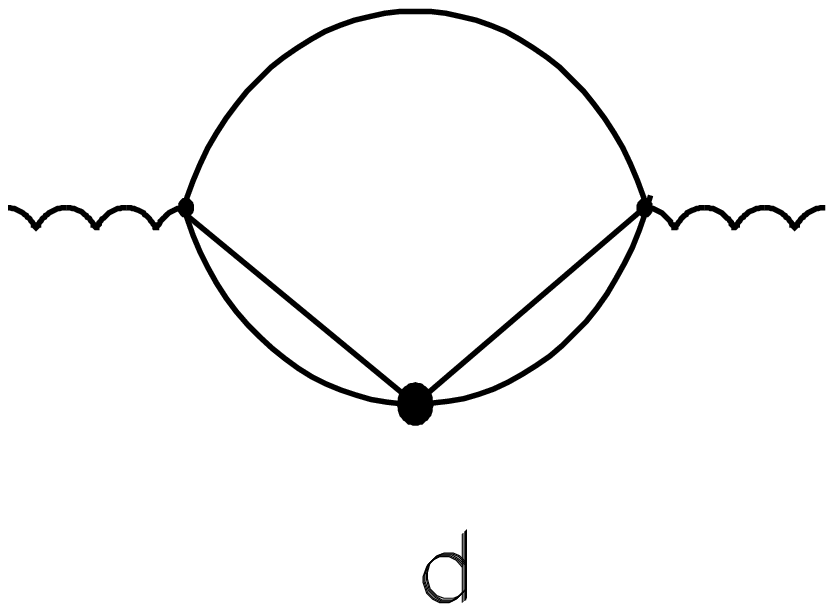,width=5cm}}
\caption{}
\end{figure}

\begin{figure}
\centerline{\epsfig{file=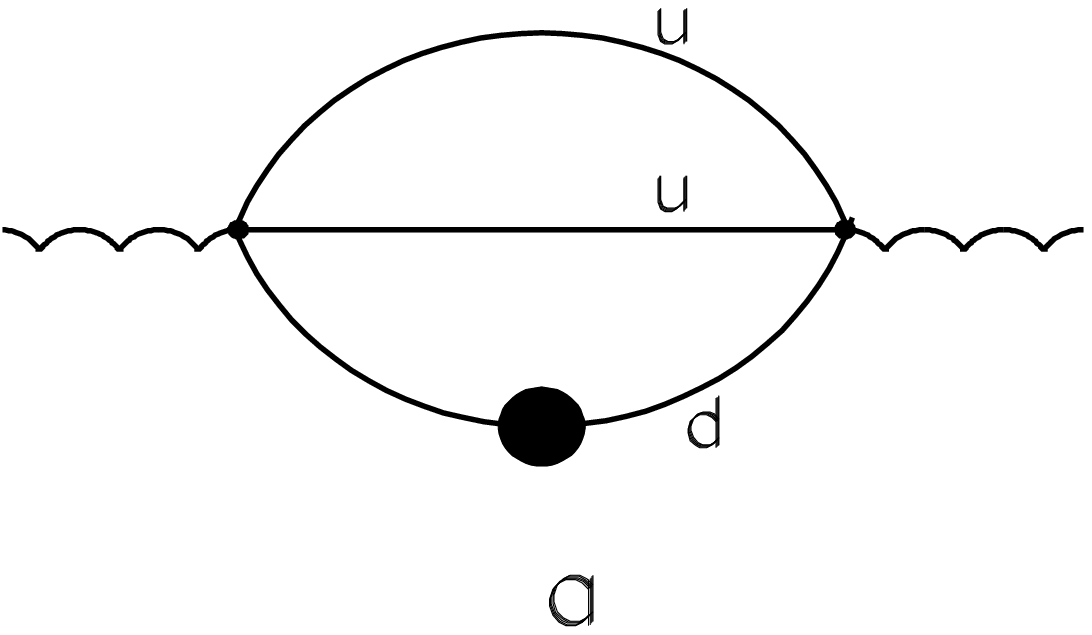,width=5cm}\epsfig{file=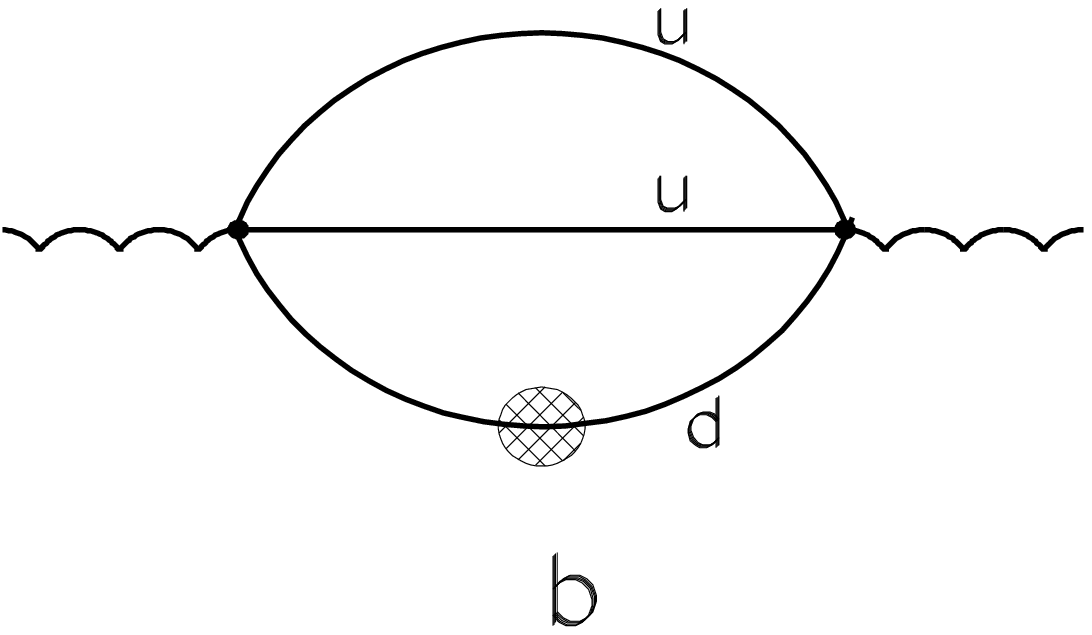,width=5cm}}
\centerline{\epsfig{file=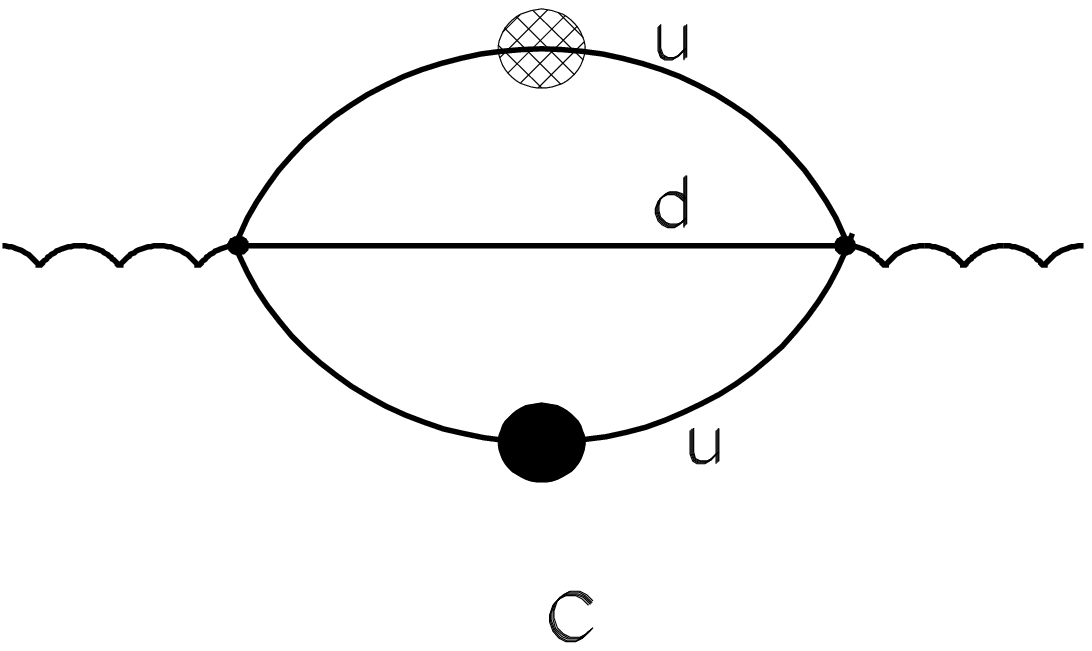,width=5cm}\epsfig{file=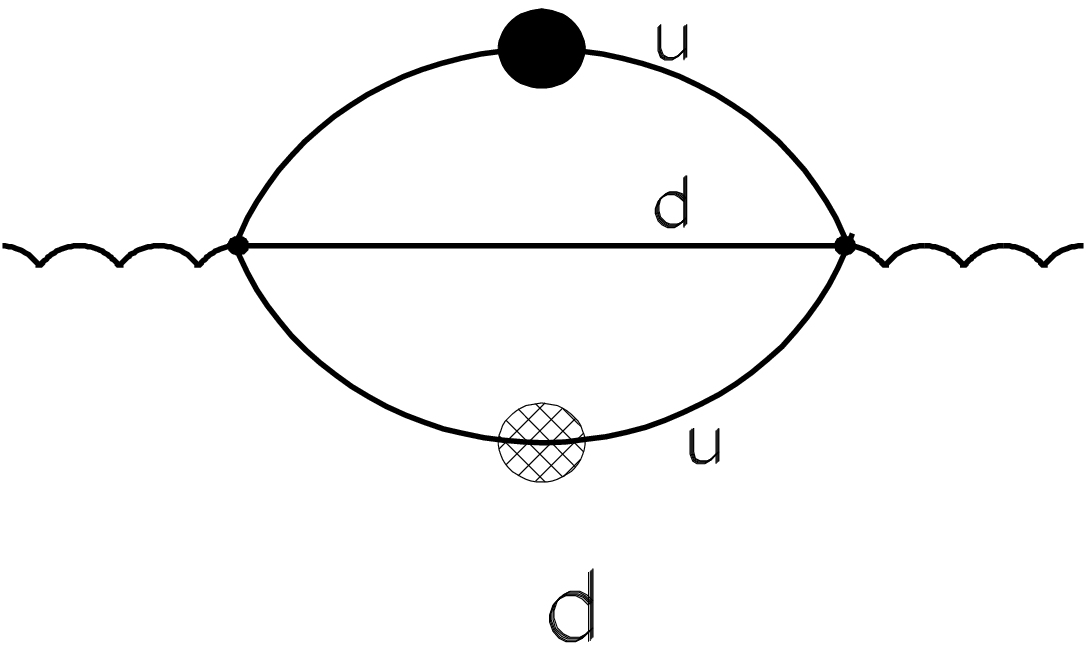,width=5cm}}
\caption{}
\end{figure}

\begin{figure}[ht] 
\centerline{\epsfig{file=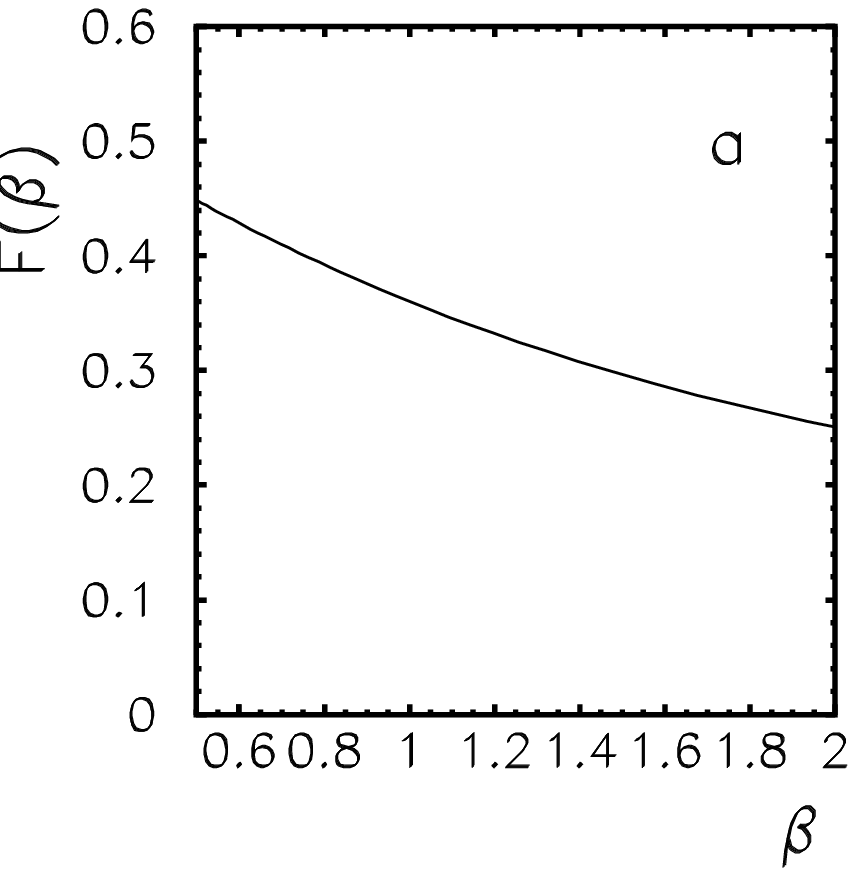,width=5.5cm}\epsfig{file=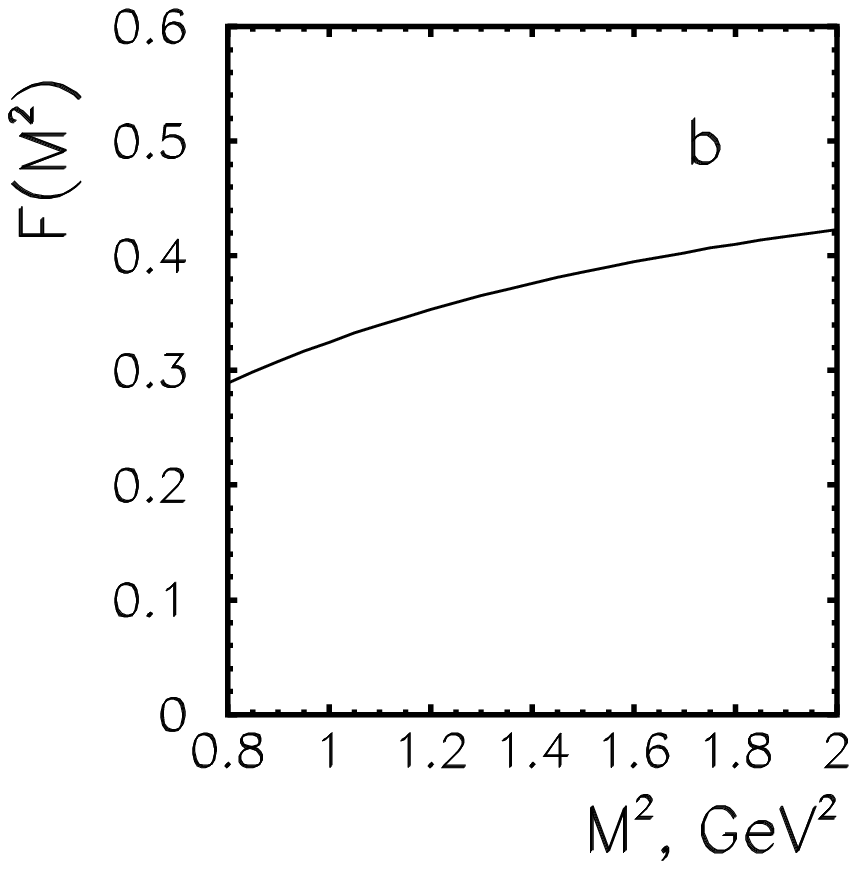,width=5.5cm}}
 \caption{}\end{figure}

\begin{figure}
\centerline{\epsfig{file=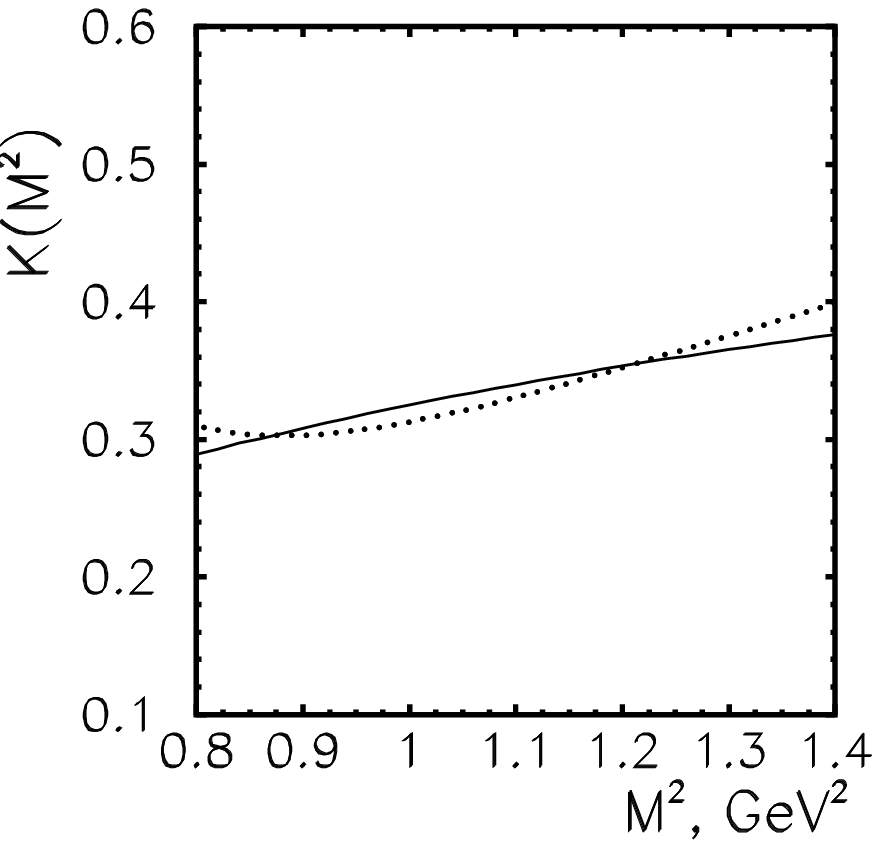,width=6cm}}
\caption{}
\end{figure}

\begin{figure}
\centerline{\epsfig{file=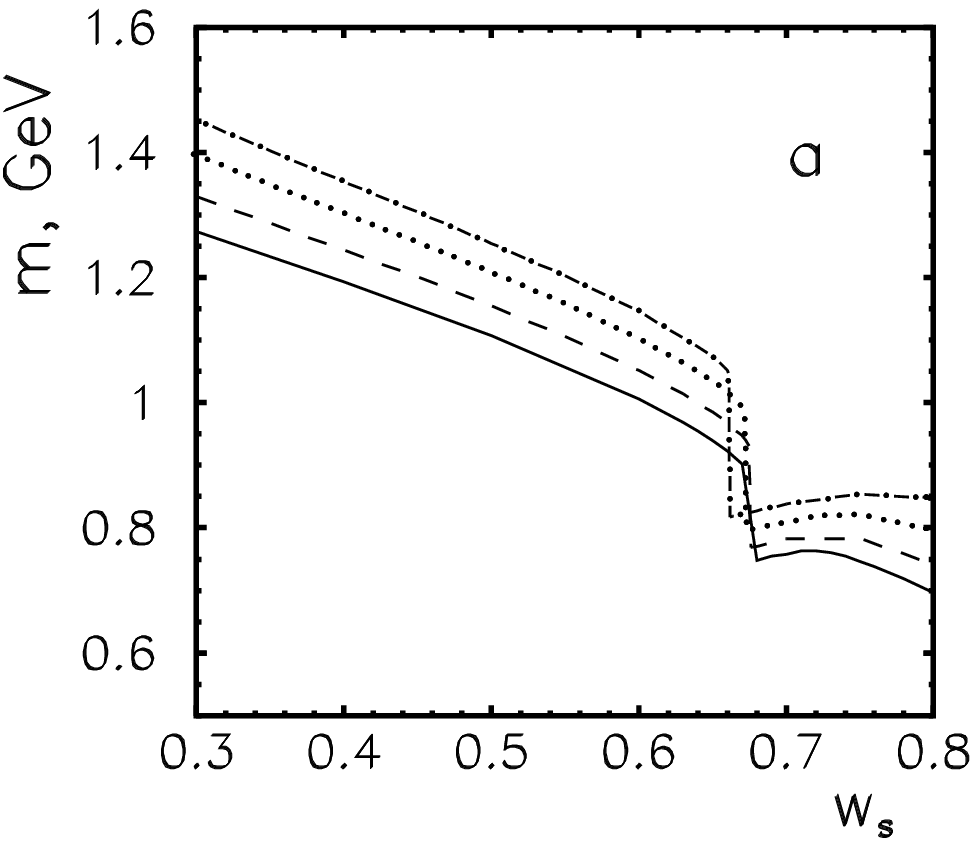,width=6cm}\epsfig{file=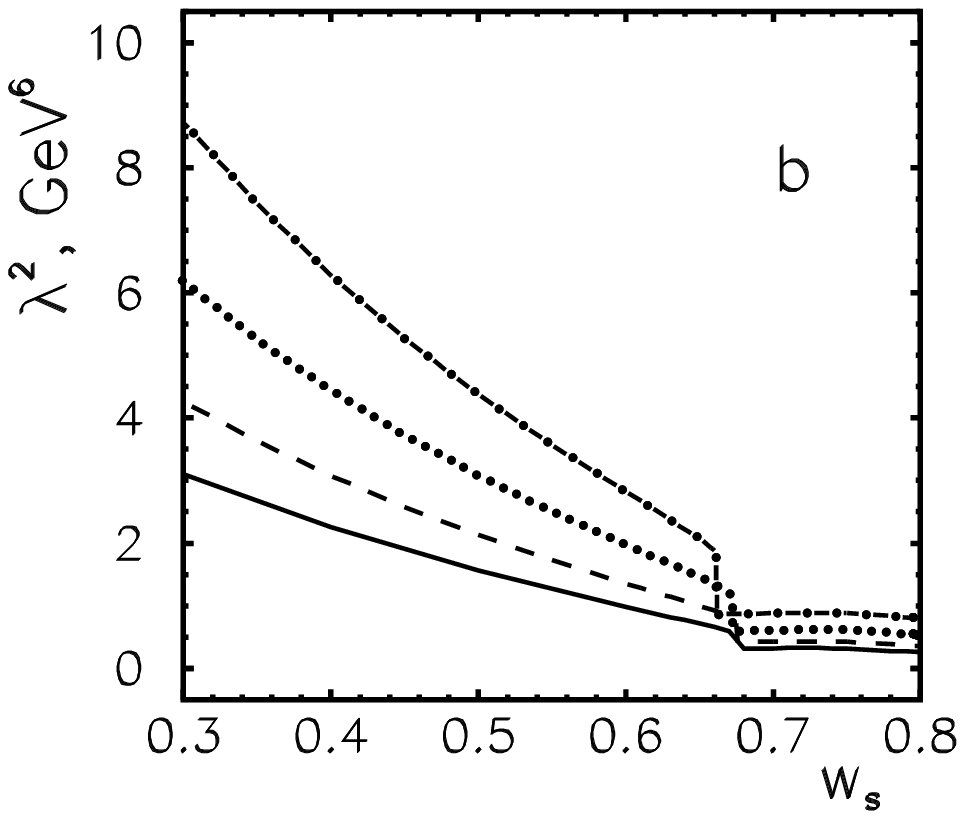,width=6cm}}
\centerline{\epsfig{file=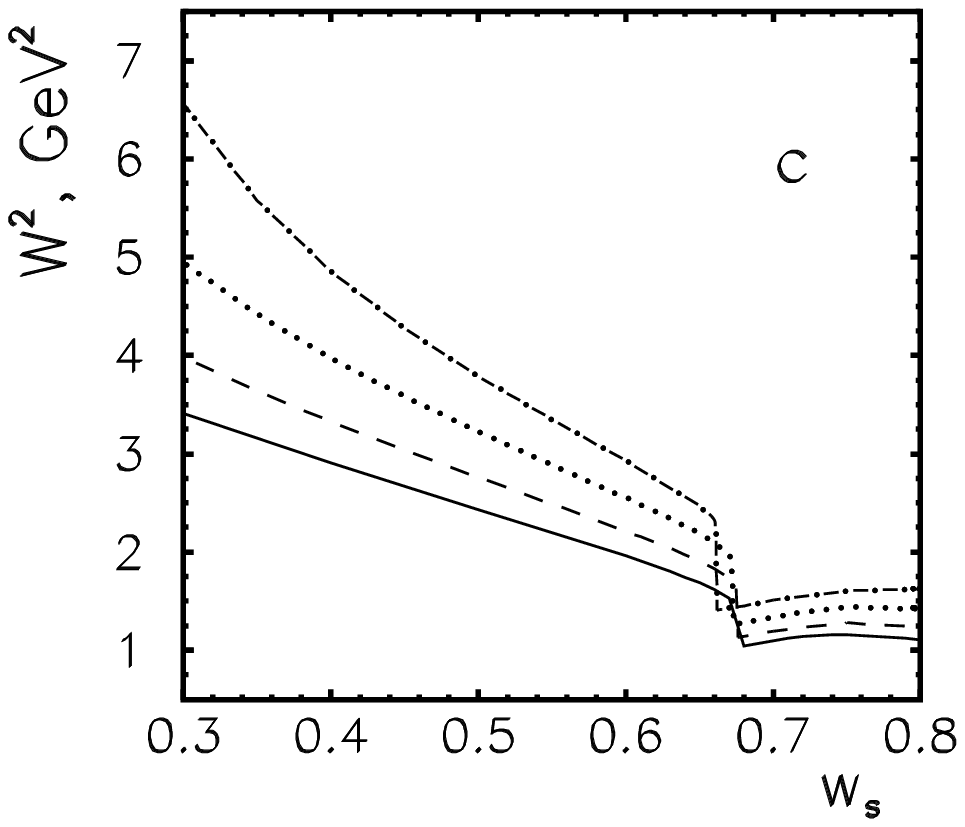,width=6cm}}
\caption{}
\end{figure}

\begin{figure}
\centerline{\epsfig{file=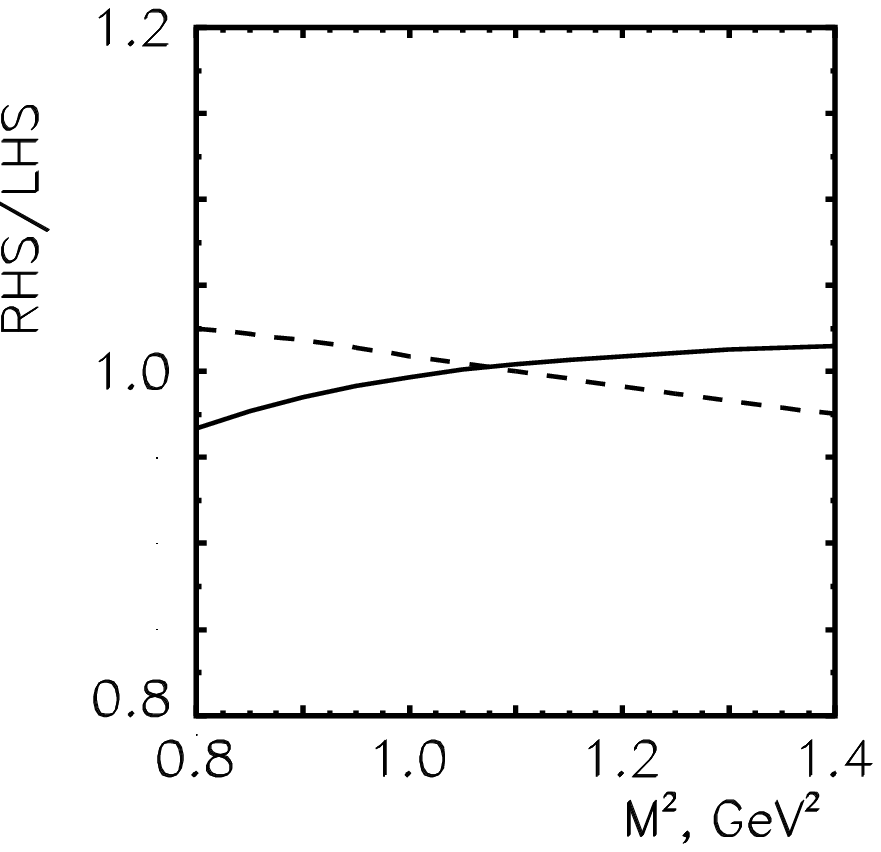,width=6cm}}
\caption{}
\end{figure}

\begin{figure}
\centerline{\epsfig{file=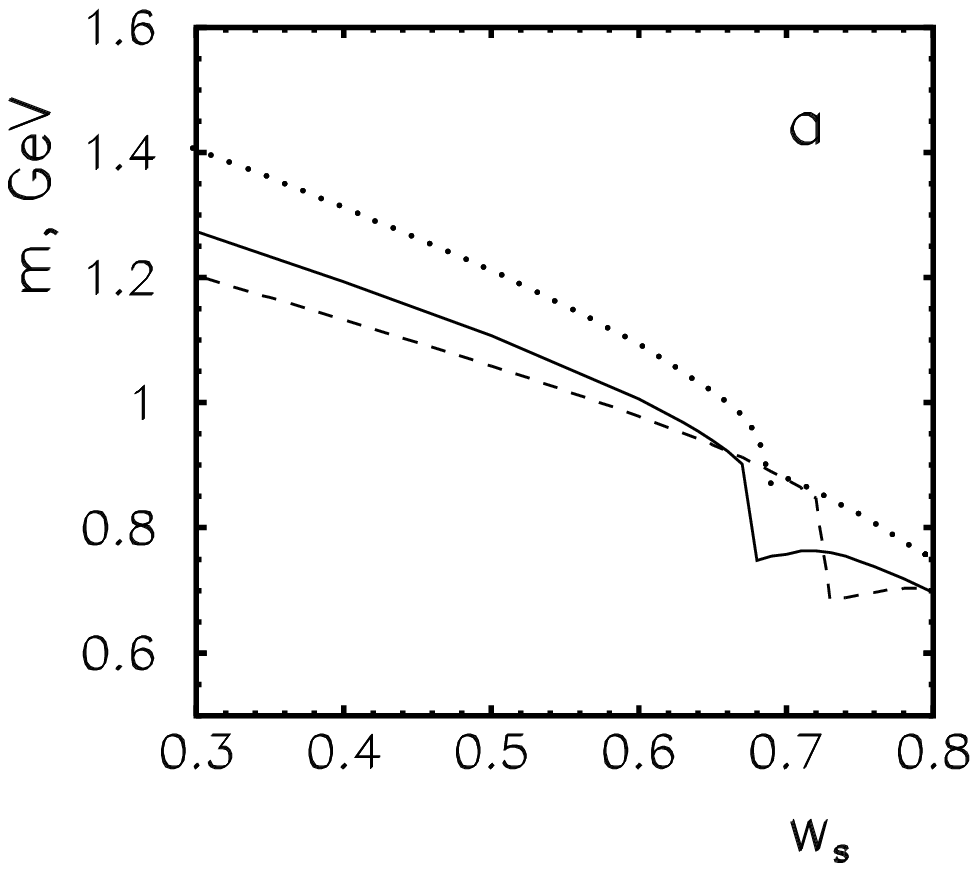,width=6cm}\epsfig{file=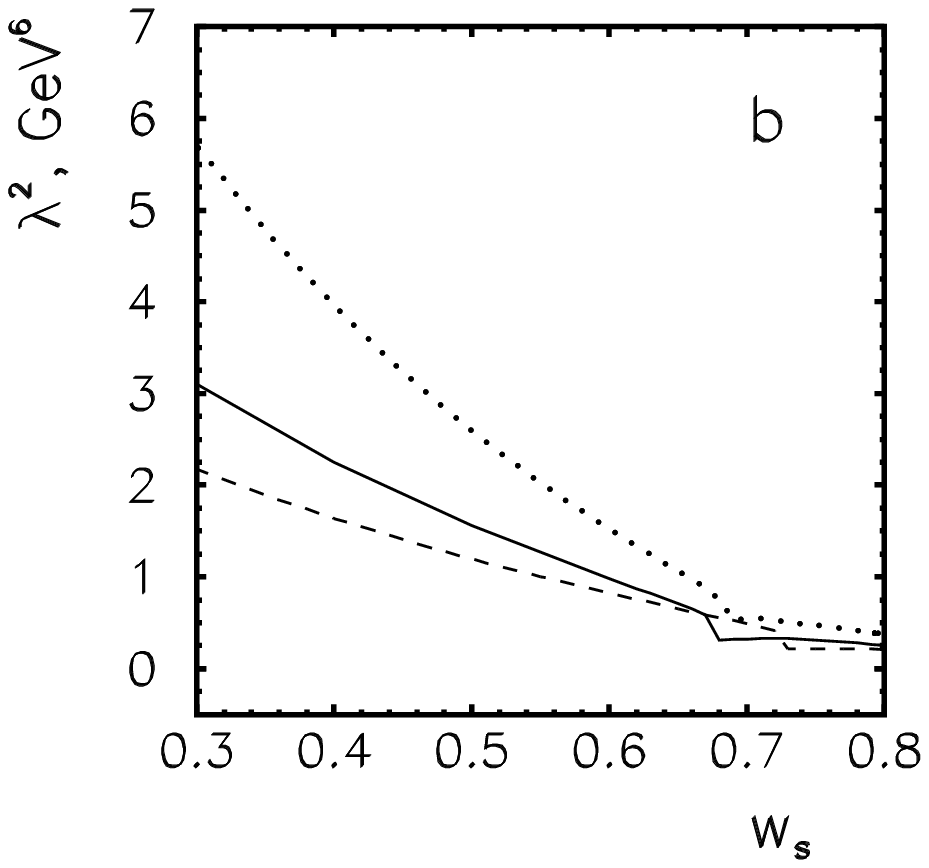,width=6cm}}
\centerline{\epsfig{file=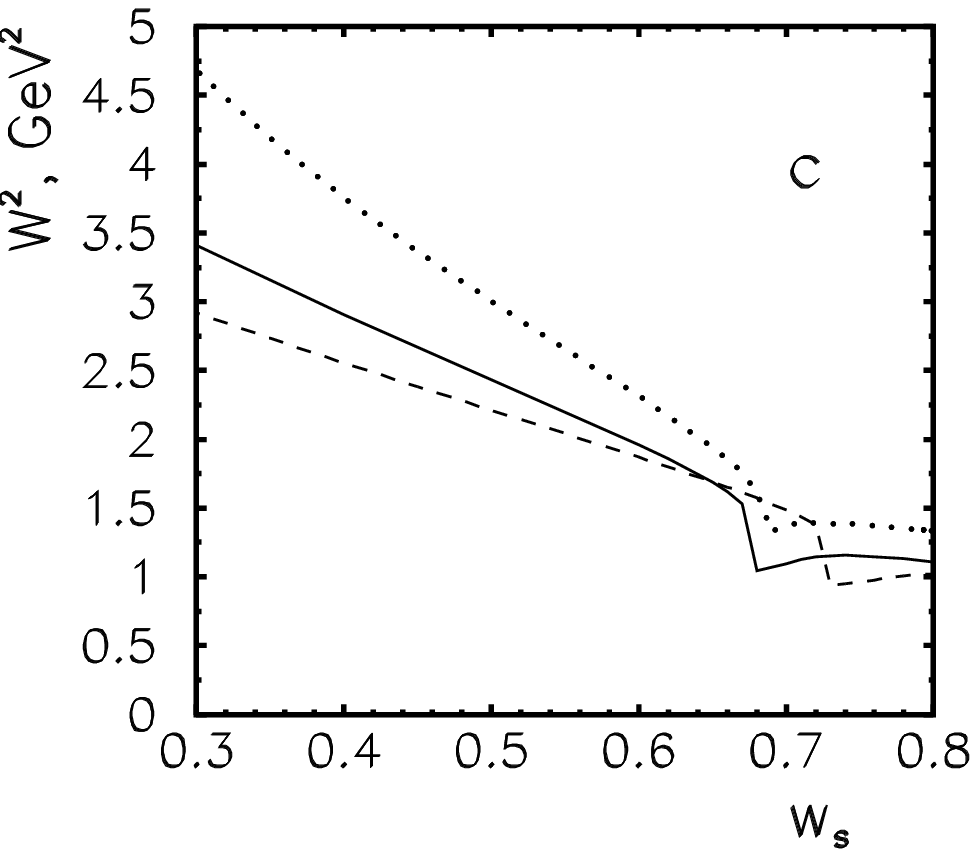,width=6cm}}
\caption{}
\end{figure}


\clearpage

\newpage

\begin{thebibliography}{}

\bibitem{1}M.A. Shifman, A.I. Vainshtein and V.I.~Zakharov, Nucl.
Phys. ~ B~{\bf147}, 385; 448; 519 (1979).

\bibitem{2} B.L. Ioffe, Nucl. Phys. ~B~{\bf188}, 317 (1981); B~{\bf191}, 591(E) (1981).

\bibitem{3} B.L. Ioffe, L.N. Lipatov and V.S.~Fadin,  Quantum
Chromodynamics, Campridge Univ. Press (2010).
\bibitem{4} K.G. Wilson, Phys. Rev. {\bf 179}, 1499 (1969).
\bibitem{3a} A.V. Radyushkin, CEBAF-TH-94-13, hep-ph/9406237; A.P. Bakulev and S.V. Mikhailov,
Phys. Rev.~D~{\bf 65}, 114511 (2002).

\bibitem{9} B.L. Ioffe, Z. Phys. C {\bf18}, 67 (1983)

\bibitem{5}E.V. Shuryak,  The QCD Vacuum, Hadrons and the Superdense Matter, World Scientific Pub. Co, Singapore (1988).

\bibitem{6} A. Ringwald, F. Schremmp, Phys. Lett. B~{\bf 459}, 249 (1999).
\bibitem{6a} H. Forkel, Phys. Rev. D~{\bf 71}, 054008 (2005).
\bibitem{7}  D.I. Dyakonov and V.Yu. Petrov, Sov. Phys. ZhETP, {\bf 89}, 361 (1985).
\bibitem{8}  D.I. Dyakonov and V.Yu. Petrov, Nucl. Phys. B {\bf 272}, 457 (1986).

\bibitem{tH} G.'t Hooft, Phys. Rev. Lett {\bf 37}, 8 (1976); Phys.
Rev.~D~{\bf 14}, 3432 (1976).
\bibitem{IR} B.L. Ioffe, Prog. Part. Nucl. Phys, {\bf 56}, 232 (2006).
\bibitem{F1} V.A. Novikov, M.A. Shifman, A.I.~Vainshtein and V.I.~Zakharov, Nucl.
Phys. ~B~{\bf174}, 378 (1980).
\bibitem{F2} H. Forkel, M.K. Banerjee, Phys. Rev. Lett. {\bf71}, 484
(1993).
\bibitem {I1} B.L. Ioffe and A.V. Smilga, Nucl.Phys.~B~{\bf 232},109 (1984).
\bibitem{10} V.A.~Sadovnikova, E.G.~Drukarev and M.G.~Ryskin, Phys.
Rev.~D~{\bf 72}, 114015 (2005).
\bibitem{M1} M.G. Ryskin, E.G. Drukarev, V.A. Sadovnikova, Phys.Atom.Nucl., accepted for publication.
\bibitem{11a} E.G.~Drukarev, M.G.~Ryskin, and V.A.~Sadovnikova, Phys.
Rev. D~{\bf 80}, 014008 (2009).
\bibitem{Shur} E.V. Shuryak, Nucl. Phys.~B~{\bf 328}, 85 (1989).
\bibitem{1a} Y. Chung, H.G.~Dosch, M.~Kremer, D.~Schall, Z. Phys.~C~{\bf 25}, 151 (1984).

\end{thebibliography}
\end{document}